\documentstyle[prd,aps,floats,psfig,amsfonts,amsbsy,amssymb]{revtex}
\newcommand{\M}{\mathbb}

\newcommand{\bi}{{\mathbf{i}}}
\newcommand{\bj}{{\mathbf{j}}}

\newcommand{\ti}{\textit}
\newcommand{\f}{\frac}

\newcommand{\bb}{\bibitem}
\newcommand{\BF}{\begin{figure}\begin{center}}
\newcommand{\EF}{\end{center}\end{figure}}
\newcommand{\BE}{\begin{equation}}
\newcommand{\EE}{\end{equation}}
\newcommand{\BEA}{\begin{eqnarray}}
\newcommand{\EEA}{\end{eqnarray}}
\tighten
\draft

\begin{document}

\title{Numerical Study of Length Spectra and Low-lying Eigenvalue 
Spectra of Compact Hyperbolic 3-manifolds}
\author{Kaiki Taro Inoue}
\address{Yukawa Institute for Theoretical Physics, Kyoto University,
Kyoto 606-8502, Japan}
\date{\today}

\maketitle

\begin{abstract}

In this paper,  we numerically investigate the length spectra and 
the low-lying eigenvalue spectra 
of the Laplace-Beltrami operator 
for a large number of small compact(closed) hyperbolic
(CH) 3-manifolds. The first non-zero eigenvalues 
have been successfully computed using the periodic orbit sum method, 
which are compared with various geometric quantities such as 
volume, diameter and length of 
the shortest periodic geodesic of the manifolds. 
The deviation of low-lying eigenvalue spectra of manifolds 
converging to a cusped hyperbolic manifold from the asymptotic
distribution has been measured by $\zeta-$ 
function and spectral distance.
\end{abstract}

\begin{picture}(0,0)
\put(440,200){
YITP-00-59}
\end{picture}

%%%%%%%%%%%%%%%% INTRODUCTION %%%%%%%%%%%%%%%%%%%
\section{INTRODUCTION}
Eigenmodes of the Laplace-Beltrami operator $\Delta$
on a Riemannian manifold (or orbifold)
which carry information of both local geometry and global topology
\BE
(\Delta + E)u_E=0
\label{eq:Helmholtz}
\EE
play a significant role in various kinds of physical systems.
\\
\indent
In the cosmological perturbation theory, one can interpret 
$E^{1/2}$ as
the wavenumber $k$ of an eigenfunction $u_k$ which 
characterises the scale of the
metric or matter perturbation. 
In order to distinguish models with equivalent
local geometry but non-equivalent global topology, it is crucial to
study the behavior of low-lying eigenmodes. For spatially compact models, 
the lowest non-zero wavenumber $k_1$ (the 
first ``excited'' state)  determines the maximum
fluctuation scale of the perturbation since fluctuations
on scale less than $k_1^{-1}$ are strongly suppressed.
In compact flat 3-manifolds, $k_1^{-1}$ is approximately equal to 
an inverse of the diameter(the maximum distance between two points) 
of the space. Therefore, if the space elongates in one dimension
and shrinks in the remaining dimension keeping the volume constant,
$k_1$ converges to zero. 
For compact hyperbolic(CH) 2-manifolds where one can 
deform the space continuously 
(without topology change) Rayleight's theorem
implies that $k_1$ can be arbitrarily 
close to zero\cite{Brooks}. However, subtlety arises in the case of 
CH 3-manifolds.
Because of the Mostow's rigidity theorem, one cannot deform the space
continuously. Instead, we have a series of manifolds 
with different topology converging to a cusped manifold with finite
volume (co-finite manifold).
For CH 3-manifolds $M$, orthonormal basis of $L^2(M)$ 
may include supercurvarture modes $k\!<\!1$ whereas there are not any
such modes in the simply-connected hyperbolic 3-space ${{\M H}^3}$.
If any supercurvature modes were present, the fluctuation property
on large scales would be drastically altered.  
\\
\indent
In quantum mechanics, $u_E$ can be interpreted as a wave function 
of a free particle at a stationary state with energy $E$.  The statistical
property of the energy eigenvalue $E$ and the eigenfunction $u_E$
have been investigated for exploring the imprints of 
classical chaos in the corresponding quantum system(e.g. see\cite{Boh}
 and other articles therein).
Because any classical dynamical systems of a free particle 
in CH spaces are strongly chaotic (K-systems), the semiclassical 
behavior of the statistical property of eigenvalues and eigenfunctions 
in these spaces has been intensively studied\cite
{Balazs,AS1,AS2,AS3,BSS,AS4,AS5,AM}. 
It is the Gutzwiller trace
formula\cite{Gutzwiller} that relates a set of periodic orbits(=geodesics)
to a set of energy eigenstates and gives the semiclassical
correspondence for classically chaotic systems.
Interestingly, for CH spaces,
the Gutzwiller trace formula gives an \textit{exact} relation which had 
been known as the Selberg trace formula in mathematical
literature\cite{Selberg}. 
The trace formula gives an alternative method to compute the
eigenvalues in terms of periodic orbits. The poles of energy
Green's function are generated as a result of interference of 
waves each one of which corresponds to a periodic orbit. 
Roughly speaking, periodic 
orbits with shorter length contribute to the deviation from the
asymptotic eigenvalue distribution on larger energy scales. In fact,
zero-length orbits produce Weyl's
asymptotic formula. 
Because periodic orbits can be obtained algebraically, the periodic orbit
sum method enables
one to compute low-lying eigenvalues for a large sample of manifolds
or orbifolds systematically if each fundamental group is known beforehand.
The method has been used
to obtain eigenvalues of the Laplace-Beltrami operator on 
CH 2-spaces and a non-arithmetic 3-orbifold\cite{AS1,AS3,AM}. 
However, it has not been applied to any CH
3-manifolds so far.
\\
\indent
To date, various numerical techniques have been applied
to the eigenvalue problems for solving the Helmholtz 
equation (\ref{eq:Helmholtz})
with periodic boundary conditions (manifold
case), Neumann and Dirichlet boundary conditions(orbifold case).
Eigenvalues of
CH 2-spaces has been numerically obtained
by many authors\cite{Balazs,AS1,AS2,BSS,AS4,Hejhal}. Eigenvalues of
cusped arithmetic and cusped non-arithmetic 3-manifolds with finite volume 
have been obtained by Grunewald
and Huntebrinker using a
finite element method\cite{Grunewald}.  Aurich and Marklof have computed
eigenvalues of a non-arithmetic 3-orbifold using the direct boundary
element method(DBEM)\cite{AM}. To date, computation of eigenvalue spectra is
limited to a small number of 3-manifolds. The author
has succeeded in computing eigenvalues of the 
Thurston manifold, the second smallest one using the 
DBEM\cite{Inoue1}, and later the Weeks
manifold, the smallest one in the known CH manifolds using 
the same method\cite{Inoue3}. Cornish and Spergel 
have also succeeded in calculating eigenvalues of these manifolds and
10 other CH manifolds based on the Trefftz method\cite{Cornish1}.   
\\
\indent
In this paper, we study the
length spectra and the low-lying eigenvalue spectra of the
Laplace-Beltrami operator on a relatively large number of small CH
manifolds which are obtained by a computer program 
``SnapPea'' by Weeks\cite{SnapPea}.
We analyse the 
fluctuating property of length spectra and
check the accuracy of the first non-zero eigenvalues obtained by the 
periodic orbit sum method based on the trace formula. 
We also relate the low-lying eigenvalues to diameter, volume and length
of the shortest periodic orbit and characterise the deviation 
of the spectrum from the asymptotic distribution.
In Sec. II we briefly
describe fundamental aspects of CH 3-manifolds which we will study. 
In Sec. III we study the length spectra of CH 3-manifolds, especially
we put an emphasis on their fluctuating property. 
In Sec. IV we derive an explicit form for computing the spectral staircase 
in terms of length spectra from the trace formula.
In Sec. V we analyse the relation between the low-
lying eigenvalues and several diffeomorphism-invariant geometric 
quantities, namely, volume, diameter and length of the 
shortest periodic orbit.
 In Sec. VI the deviation of the low-lying eigenvalue spectra 
from the asymptotic distribution for manifolds that have a region
similar to the neighbourhood of a cusped point  
is measured by $\zeta-$ function and the spectral distance.
\section{Hyperbolic manifolds}
The discrete
subgroup $\Gamma$ of $PSL(2,{\M C})$ which is the orientation-preserving 
isometry group of the simply-connected hyperbolic 3-space ${\M H}^3$
is called the Kleinian group. Any CH 3-spaces (either manifold or
orbifold) can be described 
as compact quotients ${\cal M}={\M
H}^3/ \Gamma$. If we represent ${\M H}^3$ as an upper half space
($x_1,x_2,x_3$), the metric is written as
\BE
ds^2=\f{R^2(dx_1^2+dx_2^2+dx_3^2)}{x_3^2},
\EE
where $R$ is the curvature radius. In what follows,
$R$ is set to unity without loss of generality.
If we represent a point $p$ on the upper-half space, as a quaternion whose
fourth component equals zero, then the actions of  $PSL(2,{\M
C})$ on ${\M H}^3 \cup {\M C} \cup\{ \infty\} $ take the form  
\begin{equation}
\gamma: p\rightarrow p'=\f{a p+ b}{c p+ d}, 
~~~~~ad-bc=1,~~~p\equiv z + x_3 {\bj}, ~~~z=x_1+x_2 {\bi},
\end{equation}
where a, b, c and d are complex numbers and $1$, $\bi$ and $\bj$ are
represented by matrices,
\BE
1=
\left(
\begin{array}{@{\,}cc@{\,}}
1 & 0\\ 
0 & 1 
\end{array}
\right),~~~
\bi=
\left(
\begin{array}{@{\,}cc@{\,}}
i & 0\\ 
0 & -i 
\end{array}
\right),~~~
\bj=
\left(
\begin{array}{@{\,}cc@{\,}}
0 & 1\\ 
-1& 0 
\end{array}
\right).
\EE
The action $\gamma$ is
explicitly written as   
\begin{eqnarray}
\gamma:{\M H}^3\cup{\M C}\cup\{ \infty \} &\rightarrow& 
{\M H}^3\cup{\M C}\cup\{ \infty \},
\nonumber
\\
\nonumber
\\
\gamma:(z(x_1,x_2),x_3)~~ &\rightarrow& 
\Biggl
( \f{(az+b)(\overline{cz+d})+a\bar{c}x_3^2}{|cz+d|^2+|c|^2x_3^2},
\f{x_3}{|cz+d|^2+|c|^2x_3^2}\Biggr), 
\end{eqnarray}
where a bar denotes a complex conjugate.
Elements of $\Gamma$ for orientable CH manifolds
are conjugate to 
\BE
\pm \left(
\begin{array}{@{\,}cc@{\,}}
\exp({l/2+i \phi/2}) & 0\\ 
0 & \exp({-l/2-i \phi/2})
\end{array}
\right )
\EE
which are called \ti{loxodromic} if $\phi\ne0$ and
\ti{hyperbolic} if $\phi=0$.
\\
\indent
Topological construction of CH manifolds starts with a cusped manifold
with finite volume $M_c$ 
obtained by gluing ideal tetrahedra. Let us consider the case where 
$M_c$ is topologically equivalent 
to the complement of a knot $K$ or link $L$(which consists of knots)
in 3-sphere ${ \M S}^3$ or some other closed 3-spaces. A surgery in which 
one removes the tubular neighborhood $N$ of $K$
whose boundary is homeomorphic to a torus, and replace $N$
by a solid torus so that a meridian\footnote{Given 
a set of generators $a$ and $b$
for the fundamental group of a torus, a closed curve 
which connects a point $x$ in the torus with $ax$ is 
called a {\ti{meridian}} and
another curve which connects a point $x$ with $bx$ is called a
{\ti{longitude}}.}
in the solid torus
goes to $(p,q)$ curve\footnote{If $C$ connects 
a point $x$ with another point $(pa+qb)x$ where $p$ and $q$ 
are co-prime integer, $C$ is called a $(p,q)$ curve.}
on $N$ is called $(p,q)$ \ti{Dehn surgery}. Except for a finite number
of cases, Dehn surgeries on $K$ always yield CH 3-manifolds which
implies that most compact 3-manifolds are hyperbolic\cite{Thurston1}.
SnapPea can perform Dehn surgeries which have
made it possible to construct a large number of samples of CH 3-manifolds.
\\
\indent
It is known that only a finite number of CH 3-manifolds 
with the same volume exist\cite{Thurston1}. Hence the volume 
plays a crucial role in describing CH 3-manifolds. The key facts are:
the volumes of CH 3-manifolds obtained by Dehn surgeries
on a cusped manifold $M_c$ are always less than the volume 
of $M_c$; CH 3-manifolds converge
to $M_c$ in the limit $|p|,|q| \rightarrow \infty$. 
\BF
\centerline{\psfig{figure=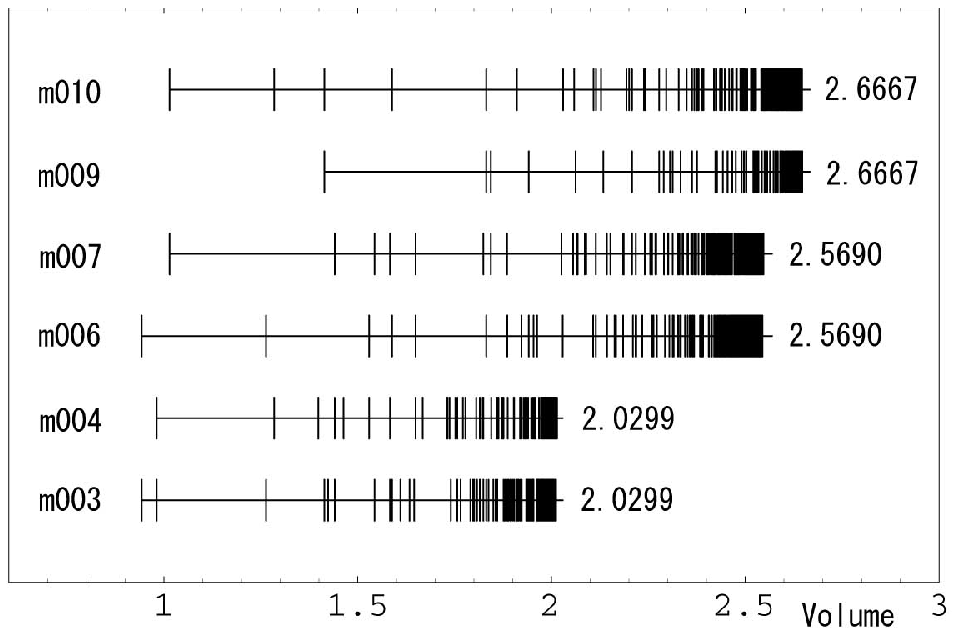,width=10cm}}
\caption{Volume spectra of CH manifolds computed by SnapPea.
A prefix ``m'' in the labelling number represents a 
cusped manifold obtained by gluing five or 
fewer ideal tetrahedra. The numbers in the right side denote the volumes 
of the corresponding cusped manifold.}
\label {fig:vol}
\EF
As shown in figure 
\ref{fig:vol}, the volume spectra are discrete but there are many 
accumulation points which correspond to the volumes of cusped
manifolds. The smallest cusped manifolds in the known manifolds
have volume $2.0299$ which are labeled as 
``m003'' and ``m004''in SnapPea. m003 and m004 are
topologically equivalent  
to the complement of a certain knot in the lens space
$L_{5,1}$ and the complement of a
``figure eight knot'' (figure \ref{fig:fig8}), respectively\cite{Matveev}.
\BF
\centerline{\psfig{figure=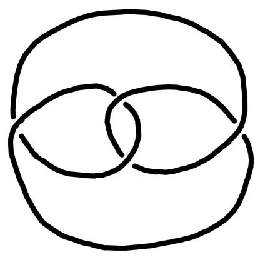,width=7cm}
\psfig{figure=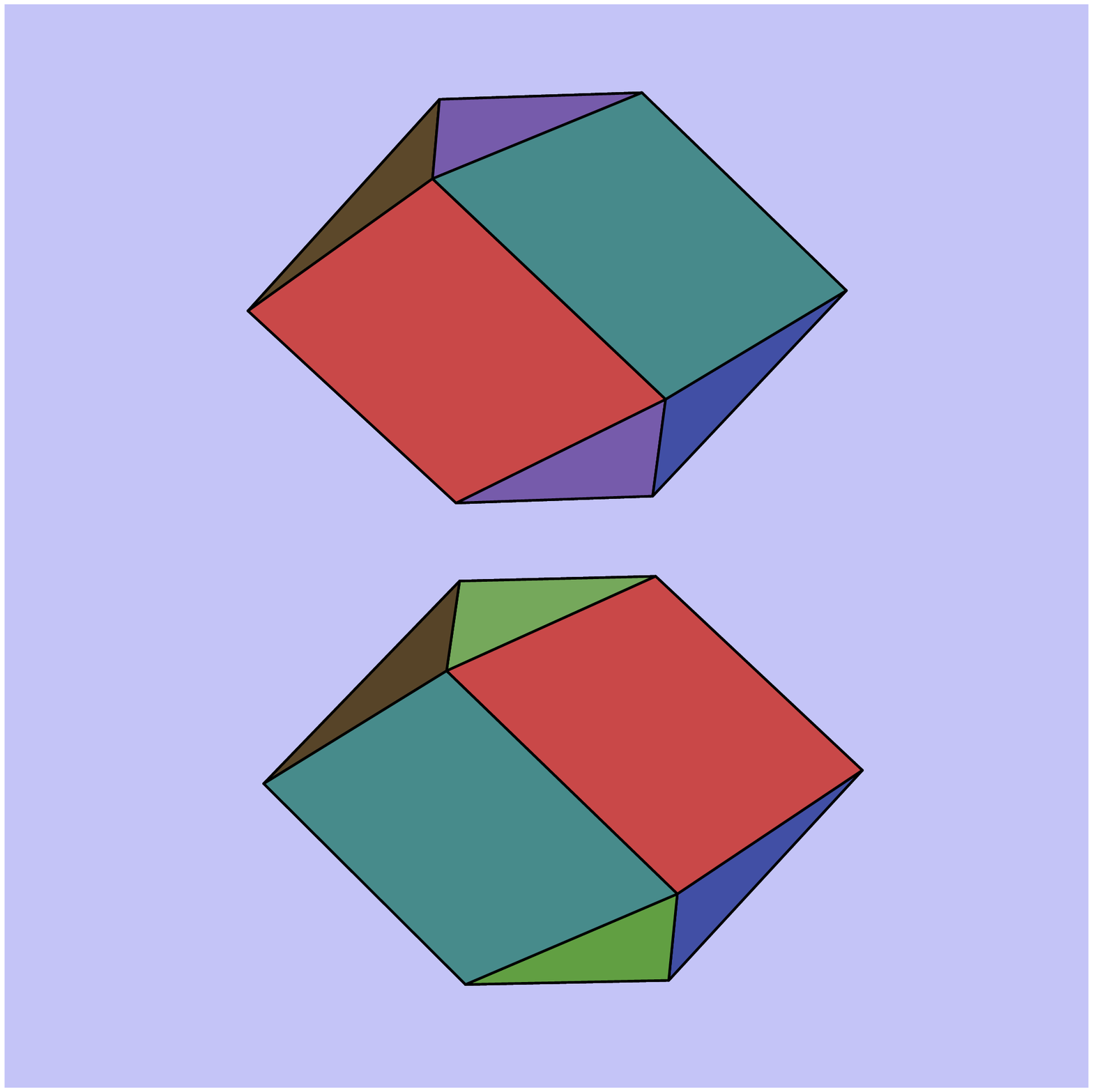,width=7cm}
}
\caption{``Figure eight knot'' (left) and
the Dirichlet domain of a cusped manifold m004
viewed from two opposite directions
in the Klein(projective) coordinates where geodesics and planes are
mapped into their Euclidean counterparts(right). 
The two vertices on the left 
and right edges of the polyhedron which are 
identified by an element of the discrete isometry group 
correspond to a cusped point. Colors on the faces correspond to the
identification maps. One obtaines the Dirichlet domain of m003 
by interchanging colors on quadrilateral faces in
the lower(or upper) right figure.}
%Given the standard coordinates 
% $(\chi,\theta,\phi)$ for
%${\M H}^3$ with curvature radius $R$ as
%$(X_0,X_1,X_2,X_3)=(R \cosh \chi,R \sinh\chi \sin\theta \cos\phi,
%R \sinh\chi \sin\theta \sin\phi, R \sinh \chi \cos \theta)$, the Klein
%coordinates are given by $(z_1,z_2,z_3)=
%(R \tanh \chi \sin \theta \cos \phi, 
%R \tanh \chi \sin \theta \sin \phi, 
%R \tanh \chi \cos \theta)$.}}
\label{fig:fig8}
\EF
(3,-1) and (-2, 3) Dehn surgeries on m003 yield
the smallest and the second smallest known manifolds, which are
called the Weeks manifold (volume=0.9427) and the Thurston
manifold (volume=0.9814), respectively. 
As $|p|$ and $|q|$ becomes large, the volumes 
converge to that of $M_c$. Similarly, one can do Dehn surgeries on
m004 or other cusped manifolds to 
obtain a different series of CH manifolds.
\section{Length Spectra}
Computation of periodic orbits (geodesics) are of crucial
importance for the semiclassical quantization of 
classically chaotic systems which will be discussed 
in the next section.   
However, in general, solving a large number of
periodic orbits often becomes an intractable problem since
the number of periodic orbits grows exponentially with an 
increase in length.
For CH manifolds periodic orbits can be calculated  
algebraically since 
each periodic orbit corresponds to a conjugacy class
of hyperbolic or loxodromic elements of the discrete isometry group 
$\Gamma$. 
The conjugacy classes can be directly computed from generators
which define the Dirichlet fundamental domain of the CH manifold
\\
\indent
Let $g_i,(i=1,...,N)$ be the generators and ${\cal{I}}$ be the
identity. In general these generators are 
not independent. 
They obey a set of relations 
\BE
\prod g_{i_1}g_{i_2}\ldots g_{i_n}={\cal{I}}, \label{eq:relation}
\EE
which describe the fundamental group of $M$.
Since all the elements of $\Gamma$ can be represented by certain
products of generators, an element $g\in \Gamma$ can be written
\BE
g=g_{i_1}g_{i_2}\ldots g_{i_n},
\EE
which may be called a ``\textit{word}'' .
Using relations, each word can be shorten to a
word with minimum length. Furthermore, all cyclic permutations
of a product of generators belonging to the same conjugacy class
can be eliminated.
Thus conjugacy classes of $\Gamma$ can be computed by generating words with 
lowest possible length to some threshold length which are reduced  
by using either relations  (\ref{eq:relation}) among the generators
or cyclic permutations of the product. 
\\
\indent
In practice, we introduce a cutoff length $l_{cut}$ 
depending on the CPU power
because the number of periodic orbits grows exponentially 
in $l$ which is a direct consequence of the exponential proliferation
of tiles (copies of the fundamental domain) in tessellation.
Although it is natural to expect a long
length for a conjugacy class described in a word with many letters,
there is no guarantee that all the the periodic orbits 
with length less than $l_{cut}$ is actually computed or not for
a certain threshold of length of the word.
\\
\indent
Suppose that each word as a transformation
that acts on the Dirichlet fundamental domain $D$. For example,
$D$ is transformed to $D'\!=\!gD$ by an element $g$.
 We can consider $gD'$s as tessellating tiles in the 
universal covering space.
If the geodesic distance $d$ between the center (basepoint) of  
$D$ and that of $gD$ is large, we can expect a long periodic orbit 
that corresponds to the conjugacy class of $g$. Tessellating tiles
to sufficiently long distance $d>l_{cut}$ makes it 
possible to compute the complex primitive
length spectra \{$L_j=l_j \exp (i \phi_j),|l_j<l_{cut}$\} where
$l_j$ is the real length of the periodic orbit of a conjugacy class
with one winding number and $\phi_j$ is the phase of the 
corresponding transformation. We also compute multiplicity number
$m(l_j)$ which counts the number of orbits having the same $l_j$ 
and $\phi_j$.  
\\
\indent
In general, the lower limit of the
distance $d$ for computing a complete set of length spectrum 
for a fixed $l_{cut}$ is not known 
but the following fact has been proved by
Hodgson and Weeks\cite{HW}.
In order to compute a length spectra of a CH 
3-manifold (or 3-orbifold) with length less than $l$,
it suffices to compute elements $\{g\}$ satisfying
\BE
d(x,gx)< 2 \cosh^{-1} (\cosh R \cosh l/2),
\label{eq:sufficient}
\EE
where $x$ is a basepoint
and $R$ is the spine radius\footnote{Spine radius $R$ is equal to the
maximum over all the Dirichlet fundamental domain's edges of the 
minimum distance from the edge to the basepoint. Note that R is finite 
even for a cusped co-finite manifold.}. Note that there is a unique
geodesic which lies on an invariant axis for each hyperbolic
or loxodromic element.
SnapPea can compute
length spectra of CH spaces either by the ``rigorous method''
based on the inequality (\ref{eq:sufficient}) 
or the ``quick and dirty'' method by setting the tessellating radius 
$d$ by hand. 
I have used the former method for manifolds 
with small volume ($<1.42$), but  
the latter method ($d=l_{cut}+0.5$) has been also used for some 
manifolds with large volume ($>1.42$) since the 
tessellating radius given by the former method
is sometimes so large that the computation time becomes too long. 
The detailed algorithm is summarized in appendix A.
\\
\indent
The asymptotic behavior of the classical staircase $N(l)$ which counts 
the number of primitive 
periodic orbits with length equal to or less than $l$
for CH 3-spaces can be written in terms of $l$ and the topological entropy 
$\tau$ \cite{Margulis}
\BE
N(l)\sim \textrm{Ei}(\tau l) \sim \f{\exp({\tau l})}{\tau l}, 
~~l \rightarrow \infty.
\label{eq:asympl}
\EE
\BF
\centerline{\psfig{figure=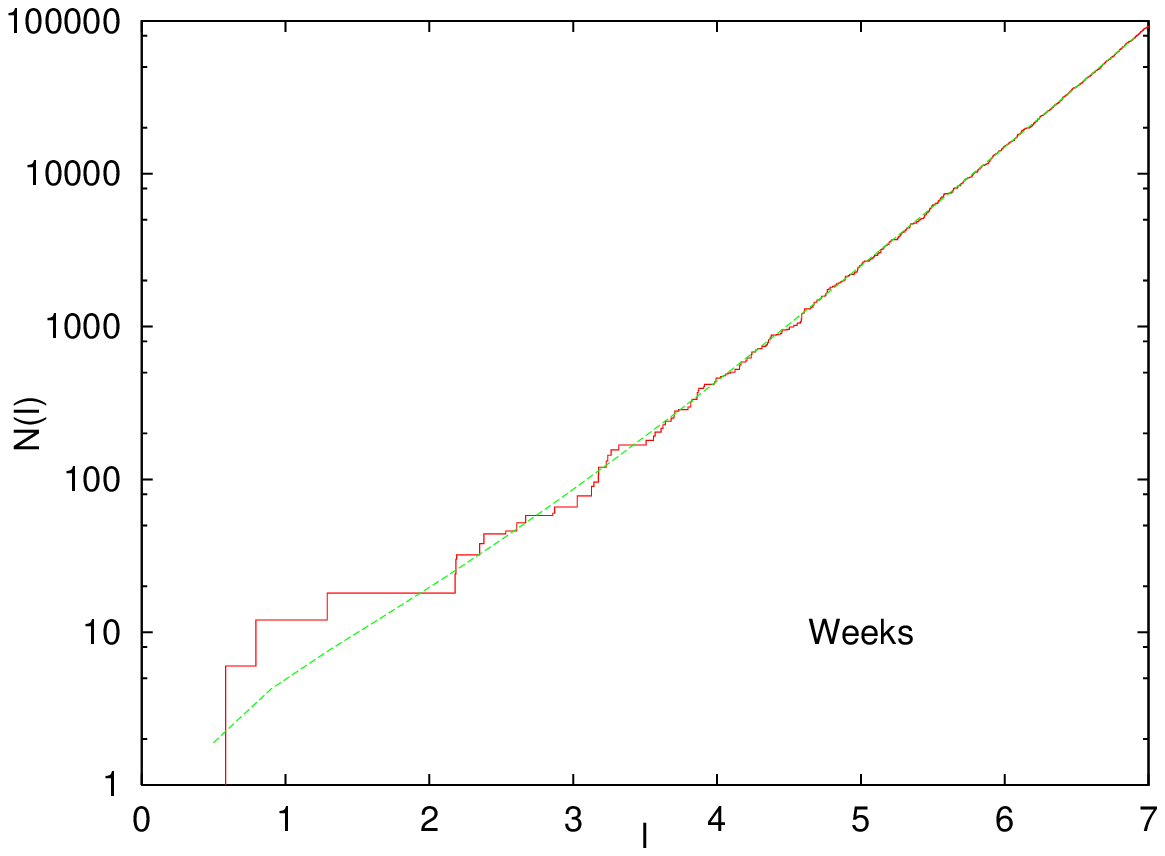,width=8cm}
\psfig{figure=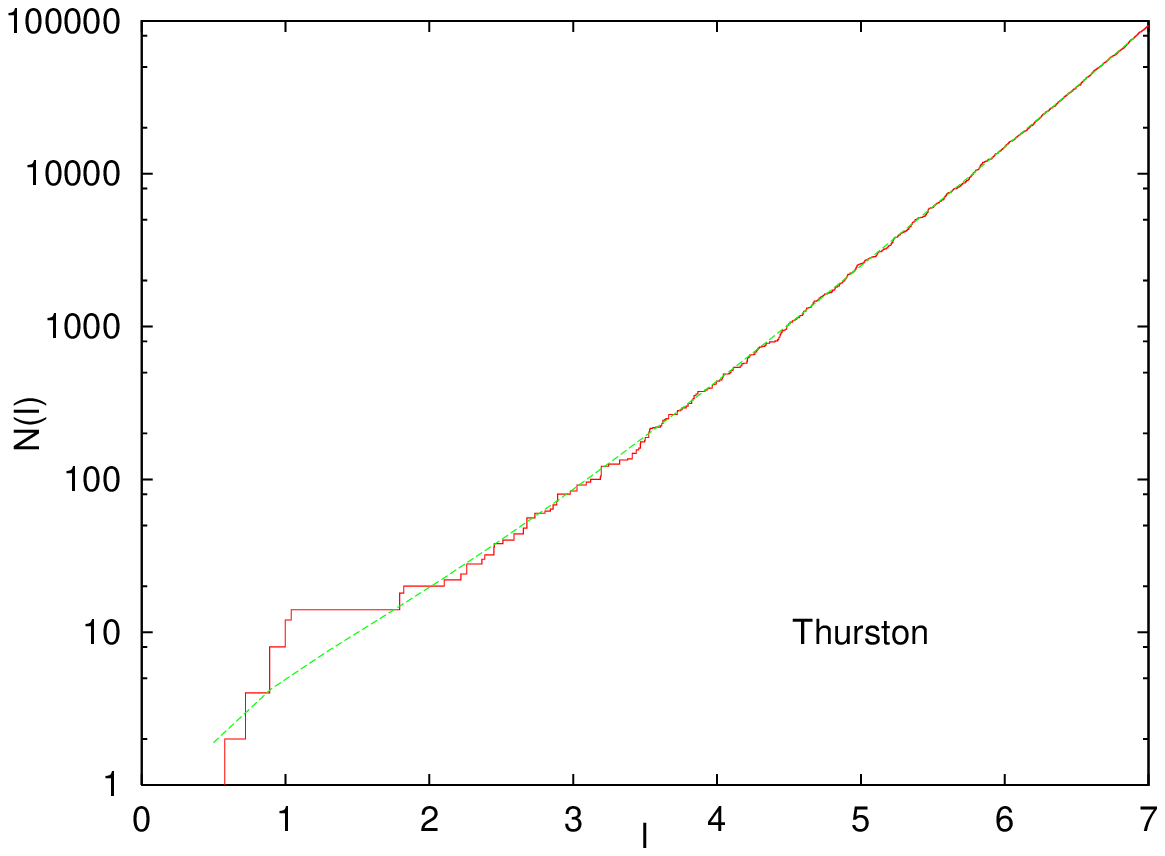,width=8cm}}
\caption{The classical staircases $N(l)$ for the 
Weeks manifold and the Thurston manifold with the asymptotic
distribution (\ref{eq:asympl}).}
\label{fig:length}
\EF
The topological entropy for $D$-dimensional CH spaces is given by 
$\tau=D-1$. A larger topological entropy implies that the efficiency
in computation of periodic orbit is much less for higher dimensional
cases\cite{AM}.
\\
\indent
In figure \ref{fig:length}, 
the computed classical staircases ($l_{cut}=7.0$) 
are compared with the asymptotic formula for the smallest (Weeks) 
manifold and the second smallest (Thurston) manifold. For both cases, an
asymptotic behavior is already observed at $l\sim3.5$. 
\\
\indent
Although the asymptotic behavior of the 
classical staircase $N(l)$ does not depend on the topology
of the manifold, the multiplicity number $m(l)$ does. 
In fact, it was Aurich and Steiner who firstly noticed that 
the locally averaged multiplicity number 
\BE
<\!m(l)\!>=\f{1}{N} \sum_{l-\Delta l/2 < l_i < l+\Delta l/2} g(l_i), 
~~~(N\!=\!\textrm{total number of terms} )
\EE 
grows exponentially $<\!m(l)\!>\sim e^{l/2}/l$
as $l\!\rightarrow\!\infty$ for
arithmetic 2-spaces (manifolds and orbifolds)\cite{AS0,BSS,ASS}. 
since the length $l$ of the periodic orbits
are determined by algebraic integers in the form
$2 \cosh(l/2)=\ti{algebraic integer}$\footnote{For a two-dimensional
space, the classical staircase has an asymptotic form
$N(l)\sim \exp(l)/l $. On the other hand, the  classical staircase for 
distinct periodic orbits has a form $\hat{N}(l)\sim \exp(l/2)$
for arithmetic systems. Because $m(l)d\hat{N}=dN$ we have $m(l)\sim
\exp(l/2)/l$ as $l \rightarrow \infty$.}.
The failure of 
application of the random matrix theory 
to some CH spaces may be attributed to the 
arithmetic property. For non-arithmetic
spaces, one expects that the multiplicities are determined
by the symmetries (elements of the isometry group) of the space.  
However, in the case of a non-arithmetic 3-orbifold,
it has been found that $<\!m(l)\!>$ grows exponentially in the form
$e^{b l}/(c l)$ where $b$ and $c$ are fitting
parameters\cite{AM}. This fact might implies that the symmetries of long 
periodic orbits are much larger than that of the space
even in the case of non-arithmetic systems.
\\
\indent
For 3-manifolds, one can expect that the property of  $<\!m(l)\!>$ 
for 3-orbifold also holds.
The locally averaged multiplicities $(\Delta l=0.2)$ for
the smallest twelve examples which include seven
arithmetic and five non-arithmetic 3-manifolds \cite{CGHN}
have been numerically computed using SnapPea.
\BF
\centerline{\psfig{figure=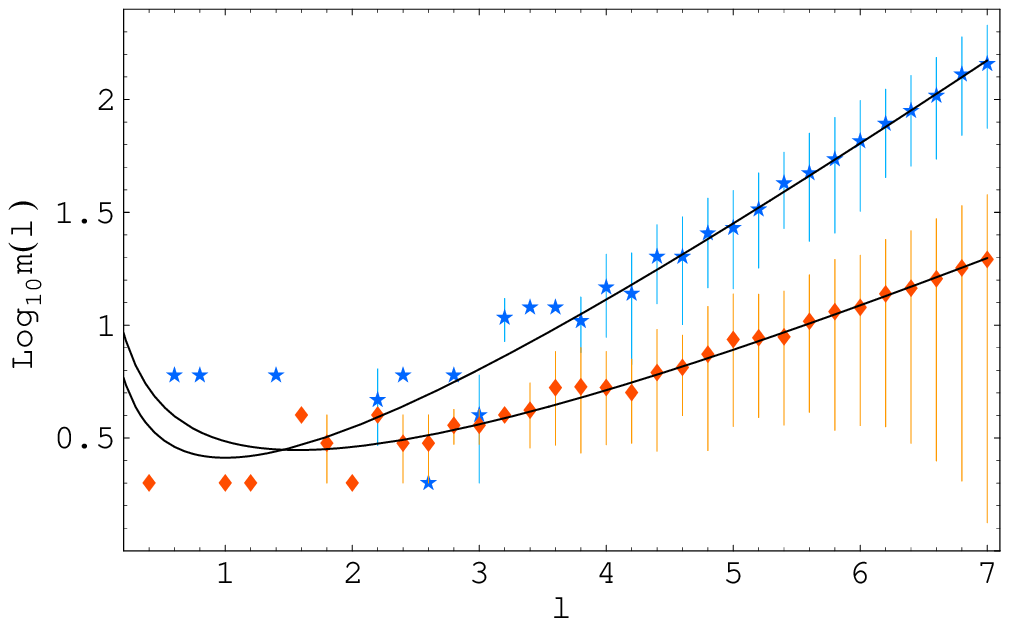,width=12cm}}
\caption{Plots of locally averaged multiplicities ($\Delta l=0.2$)
with one-sigma errors and the fitting curves
for the Weeks manifold m003(-3,1) which is
arithmetic (star) and a non-arithmetic manifold m004(1,2)
(diamond). The fitting curves ($a \exp(l)/l $ for the former and
$\exp(b l)/(c l)$ for the latter) are obtained by the least square method
using data $3.0<l<7.0$. }
\label{fig:aveg}
\EF
\begin{table}
\begin{center}
\setlength{\tabcolsep}{3pt}
\begin{tabular}{rrccrrr}   
\multicolumn{1}{c}{manifold} &
\multicolumn{1}{c}{volume} &
\multicolumn{1}{c}{A/N} &
\multicolumn{1}{c}{G} &
\multicolumn{1}{c}{$a$} &
\multicolumn{1}{c}{$b$} &
\multicolumn{1}{c}{$c$} 
\\ \hline
m003(-3,1)&0.9427   &A  &D6  &0.9514 &- &- \\ \hline  
m003(-2,3)&0.9814   &A  &D2  &0.5667 &- &- \\   \hline  
m007(3,1)&1.0149    &A  &D2$\dagger$ & 0.7108 &- &- \\ \hline  
m003(-4,3)&1.2637   &A  &D4  &0.8364  &-  &- \\ \hline  
m004(6,1)&1.2845    &A  &D2  &0.6066  &- &-  \\ \hline  
m004(1,2)&1.3985    &N  &D2  &-       &0.6360 & 0.6180\\ \hline        
m009(4,1)&1.4141    &A  &D2  &0.5362  &- &- \\ \hline
m003(-3,4)&1.4141   &A  &D2  &0.5655  &- &- \\ \hline  
m003(-4,1)&1.4236   &N  &D2  &-       &0.5933 &0.5912\\   \hline  
m003(3,2)&1.4407    &N  &D2  &-       &0.6018 &0.5532 \\ \hline  
m004(7,1)&1.4638    &N  &D2  &-       &0.5693 &0.4829\\ \hline  
m004(5,2)&1.5295    &N  &D2  &-       &0.5780 &0.5590\\ \hline  
\end{tabular}
\end{center} 
\label{tab:coefficients}
\caption{Coefficients of the fitting curves which describe the average
behavior of locally averaged multiplicities $<\!m(l)\!>$ for
arithmetic (A) and non-arithmetic (N) 3-manifolds. G denotes the
isometry group (symmetry group). Fitting parameters
$a,b$ and $c$ are obtained by the least square method
using data $3.0<l<7.0$. $\dagger$ For m007(3,1), the isometry group may 
be larger than D2.} 
\end{table}
From figure \ref{fig:aveg}, one can see that the  
difference in the behavior of $<\!m(l)\!>$ between the
non-arithmetic manifold and the arithmetic one is manifest. 
As observed in 3-orbifolds, averaged multiplicities behave as 
\BEA
<\!m(l)\!>&=&a \f{\exp{l}}{l},~~~~~~\textrm{(arithmetic)},
\\
&=&\f{\exp{b l}}{c l},~~~~~\textrm{(non-arithmetic)},
\EEA
where $a$ depends on the discrete isometry group while $b$ and
$c$ are fitting parameters. From table I, one observes that 
arithmetic manifolds having a larger
symmetry group have a larger value of $a$.
The growth rates for
non-arithmetic manifolds ($b\sim 0.56$) 
are always less than that for arithmetic
manifolds ($b=1$) 
but nevertheless exponential. 
\section{Periodic orbit sum method and spectral staircase}
Gutzwiller's periodic orbit theory provides a semiclassical
quantization rule for classically chaotic systems.
The theory is expressed in form of a semiclassical approximation 
($\hbar \rightarrow 0$) of the trace of the energy Green's operator
(resolvent operator) $\hat{G_E}\!=\!(\Delta+E)^{-1}$ 
in terms of the length of periodic 
orbits (geodesics) $\{ L_i \}$ which is known as the
\ti{Gutzwiller trace formula}\cite{Gutzwiller}.
For the dynamical system
of a free massive particle on a CH space known as the 
Hadamard-Gutzwiller model, the periodic orbits give 
the exact eigenvalues, and the relation is no longer 
semi-classical approximation. 
In mathematical literature, the trace formula is known as the 
{\it{Selberg trace formula}} \cite{Selberg}. 
In what follows we consider
only orientable CH 3-manifolds(denoted as CH manifolds)
(for general cases including
orbifolds, see \cite{AM}). 
The Selberg trace formula for a CH manifold
$M\!=\!{\M H}^3/ \Gamma$ ($\Gamma$ is a
discrete isometry group containing
only hyperbolic or loxodromic elements) can be written as 
\BEA
Tr(\hat{G}_E-\hat{G}_{E'})
&=&-\f{v(M)}{4 \pi i}(p-p')
\nonumber
\\
&&
-\sum_{\{g_\tau\}}
\f{l(g_{\tau_0})}{4(\cosh l(g_\tau)-\cos \phi(g_\tau))} \Biggl ( \f{\exp(-i 
p l(g_\tau))}{ip}-\f{\exp(-i p' l(g_\tau))}{ip'}\Biggr ), \label{eq:ST}
\EEA
where $p^2\!=\!E-1$, $v(M)$ 
denotes the volume of $M$, $l(g_\tau)$
is the (real) length of the periodic orbit of transformation $g_\tau
\in \Gamma$. $g_{\tau_0}$ is a transformation 
that gives the shortest length of the periodic orbit $l(g_{\tau_0})$
which commutes with $g_\tau$. $\phi(g_\tau)$ is the phase of the
transformation $g_\tau$. The sum in (\ref{eq:ST}) 
extends over $\Gamma$ conjugacy classes 
\BE
\{g_\tau\}:=\{g'_\tau|g_\tau'=h g_\tau h^{-1}, h \in \Gamma \}
\EE
of hyperbolic ($\phi\!=\!0$) or loxodromic elements ($\phi\!\neq\!0$).
However the periodic orbit sum in (\ref{eq:ST}) which is known as 
Maa\ss-Selberg series converges at only
complex energy such that $\textrm{Im}~ p\!<\!-1$ and $\textrm{Im}~p'\!<\!-1$. In order to obtain real eigenvalues, one needs to multiply
the trace by some suitable analytic ``smoothing'' function  $h(q)$
that satisfies:
\\
(i):$h(q)=h(-q)$;
\\
(ii):$h(q)={\cal{O}}(|q|^{-3-\delta})$ for $\delta>0$ as $|q|\rightarrow \infty$;
\\
(iii):$h(q)$ is analytic in the strip ${|\textrm{Im}~ q|<1+\epsilon}$ for 
$\epsilon>0$.
\\ 
Multiplying (\ref{eq:ST}) by $q~ h(q)/(\pi~ i)$ and integrating it
over $q$ from $-\infty$ to $\infty$, one obtains the \textit{general Selberg trace formula},
\BEA
\sum_{n=0}^\infty h(p_n) = &-& \f{v(M)}{2 \pi} \tilde{h}''(0)
\nonumber
\\
&+&\sum_{\{g_\tau\}} \f{l(g_{\tau_0})}{2(\cosh~l(g_\tau)-\cos~
\phi(g_\tau))}~\tilde{h}(l(g_\tau)), \label{eq:GTF}
\EEA  
where 
\BE
\tilde{h}(l)=\f{1}{2 \pi} \int_{\infty}^{\infty} dq~ h(q)~\exp(-iql),
\EE
and $\tilde{h}''(0)$ is the second derivative of $\tilde{h}$ and
$p_n$ denotes a wavenumber of the corresponding eigenmode.
The sum in (\ref{eq:GTF}) is absolutely convergent for any real
eigenvalues $E_n=p_n^2+1$. One can obtain various functions of eigenvalues
such as heat kernels and energy level densities from periodic orbits by choosing an appropriate ``smoothing''
function $h(q)$. 
\\
\indent
In order to obtain eigenvalues, a simple approach is to compute the
spectral staircase
\BE
N(E)=\sum_{n=0}^\infty \theta(E-E_n),
\EE
where $E_0\!=\!0, E_1,E_2,\cdots $ are the eigenvalues of the
Laplacian and $\theta$ is the Heaviside function\cite{AS1}. 
To explore supercurvature modes $u_E$,
 $0\leq E<1$, we choose the ``smoothing'' function of
$N(E)$ as 
\BEA
h_{p\!,\epsilon }(p')
&=&
\f{1}{2}\Biggl(1-\textrm{Erf}
\biggl( \f{E'-E}{\epsilon^2} \biggr ) \Biggr),
\nonumber
\\
&=&
\f{1}{2}\biggl(1-\textrm{Erf}
\biggl( \f{p'^2-p^2}{\epsilon^2} \biggr ) \biggr),
\EEA
which is real for $E>0$. Note that $h_{p}(p')$ satisfies all the 
conditions (i) to (iii).
By taking the limit $\epsilon \rightarrow 0$, one obtains
the spectral staircase
\BE
N(E)=\lim_{\epsilon \rightarrow 0} \sum_{n=0}^\infty h_{p\!,\epsilon }(p_n). 
\EE
\\
\indent
Let us first estimate the behavior of the trace in (\ref{eq:GTF}). 
From a straightforward calculation, the zero-length
contribution can be written as
\BEA
 \tilde{h}''_{p\!,\epsilon}(0)=&-& \f{\epsilon^3}{6~ \pi^{\f{3}{2}}}
 \exp(-p^4/\epsilon^4)
\nonumber
\\
&\times& \biggl ( \f{\sqrt{2}\pi}{4} (\Gamma(3/4))^{-1}
 F(5/4,~1/2,~p^4/\epsilon^4)
+\f{3~p^2}{2~\epsilon^2}\Gamma(3/4)F(7/4,~3/2,~p^4/\epsilon^4)
\biggr ), \label{eq:zerolength}
\EEA
where $\Gamma(x)$ is the Gamma function and $F(a,b,z)$ is the 
confluent hypergeometric function. 
For $x\!=\!p^4/\epsilon^4
\!\rightarrow\!\infty $, it is asymptotically expanded as  
\BEA
F(a,b,x)&=&\f{\Gamma(b)}{\Gamma(b-a)}e^{i \pi a}x^{-a} 
\Biggl \{ \sum_{n=0}^{R-1} 
\f{(a)_n (1+a-b)_n}{n !
}(-x)^{-n}+{\cal{O}}(x^{-R})
\Biggr \}
\nonumber
\\
&+&\f{\Gamma(b)}{\Gamma(a)}e^{x}x^{a-b} \Biggl \{
\sum_{n=0}^{S-1} \f{(b-a)_n (1-a)_n}{n !
}(x)^{-n}+{\cal{O}}(x^{-S}) \Biggr\}, \label{eq:as}
\EEA
where $(a)_m\equiv \Gamma(a+m)/\Gamma(a)$.
From (\ref{eq:zerolength}) and (\ref{eq:as}), in the lowest order,
we have the average part, 
\BE
\hat{N}(p)=\lim_{\epsilon \rightarrow 0} -\f{v(M)}{2 \pi}
\tilde{h}''_{p\!,\epsilon}(0)=\f{v(M)}{6\pi^2}|p^2|^{\f{3}{2}}, 
\label{eq:Weyl}
\EE
that gives the dominant term in the Weyl asymptotic formula 
for $p\!>>\!1$.
\\
\indent
Next, we estimate the oscillating term in (\ref{eq:GTF})
\BE
\tilde{h}_{p,\epsilon}(l)=\alpha \int_{-\infty}^{\infty} dq~ \textrm{exp}
(-i q l)~q~  \textrm{exp} \Bigl(-\f{(q^2-p^2)^2}{\epsilon^4}
\Bigr),~~~~\alpha 
\equiv \f{i}{l \epsilon^2 \pi^{3/2}}. \label{eq:longh}
\EE
In the long orbit-length limit $l\!>\!>\!1$ with $p\!>\!0$,
the integrand in Eq.(\ref{eq:longh}) oscillates so rapidly that 
the dominant contribution comes from $q\!\sim\!p,$ or $-p$ 
where $(q^2-p^2)^2/\epsilon^4$ can be approximately given as 
$-4(q-p)^2p^2/\epsilon^4$. Then (\ref{eq:longh}) can be written as
\BE
\tilde{h}_{p,\epsilon}(l)\sim \f{1}{\pi l}\sin(p
l)\textrm{exp}\biggl(\f{-\epsilon^4 l^2}{16 p^2}\biggr ). \label{eq:tilh}
\EE
Thus each periodic orbit corresponds
to a wave with wavelength $2 \pi/l$ and an amplitude
which is exponentially suppressed with an increase in $l$ or a decrease
in $p$.
For a finite subset of length spectra $l_j<l_{cut}$, 
the appropriate choice for the
smoothing scale is given by   
$\epsilon\! =\!\alpha p^{1/2}$ for $p^2\!>\!0$ where 
$\alpha$ depends on $\l_{cut}$ since for a reasonable value 
of the proportional factor $\alpha$ all the contributions
from periodic orbits with large length $l>l_{cut}$ can be negligible.
For  $p^2\!<\!0$, 
an optimal choice can be obtained from a numerical
computation of Eq.(\ref{eq:longh}) directly. Comparing the obtained 
smoothed spectral staircase with the one based on the computed ``true''
eigenvalues using the direct boundary element method (DBEM), 
it is numerically found
that for  $l_{cut}\!=\!7.0$, an appropriate smoothing scale is given by 
\BE
\epsilon(k)=\cases{
0.116k^2+0.184~k+1.2 ~~(k<1) \cr
0.832k^{1/2}+0.668 ~~(k \ge 1) \cr},
\EE
where $E=k^2=p^2+1$. The eigenvalues can be computed by searching
$E$ at which $N(E)-0.5$ becomes positive integer.
It should be emphasised that the precision of computation depends on the 
value of $l_{cut}\!$ which determines the resolution scale in the
eigenvalue spectra.
\\
\indent  
In order to get eigenvalues from the smoothed spectral staircase,
one must take into account the effect of the multiplicity 
number(degeneracy number) for each eigenvalue since the 
spectral staircase is 
smoothed on larger scales for degenerated modes. 
Therefore, the numerical accuracy becomes worse if
the eigenmode has a large multiplicity number.
Fortunately, the order of the symmetry group is not so large for a
small manifold (volume$<3$). For instance, of the twelve 
smallest examples, nine manifolds have a symmetry group with order 4.    
If one assumes
that the multiplicity number is either 1 or 2 then the 
deviation $\Delta k$ from a precise value is approximately 
given by
\BE
\f{1}{2}\Biggl(1-\textrm{Erf}
\biggl( \f{k^2-(k+\Delta k)^2}{\epsilon^2} \biggr ) \Biggr)=\f{1 
\pm 0.5}{2}.
\label{eq:deltak}
\EE
For instance, if one uses a length spectrum $l<7.0$ then  
(\ref{eq:deltak}) gives 
$\Delta k\!=\!0.30,0.33$ for $k\!=\!5.0,3.0$, respectively. 
If one permits the multiplicity number as much as $6$,
then the expected precision becomes 
$\Delta k\!=\!0.49,0.54$ for $k\!=\!5.0,3.0$, respectively.
\\
\BF
\centerline{\psfig{figure=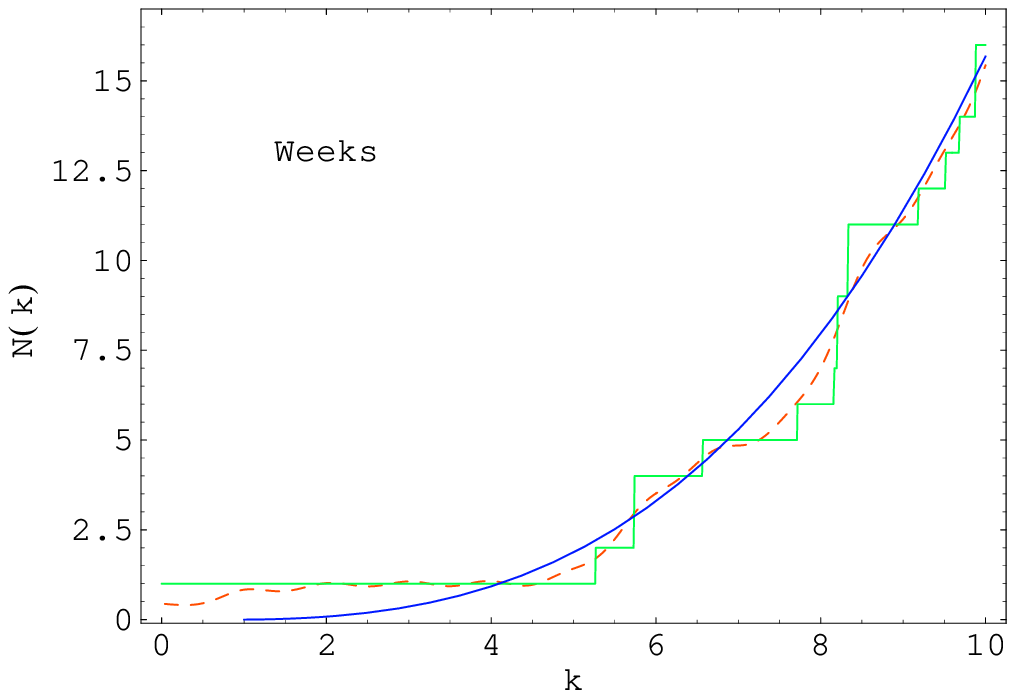,width=8cm}
\psfig{figure=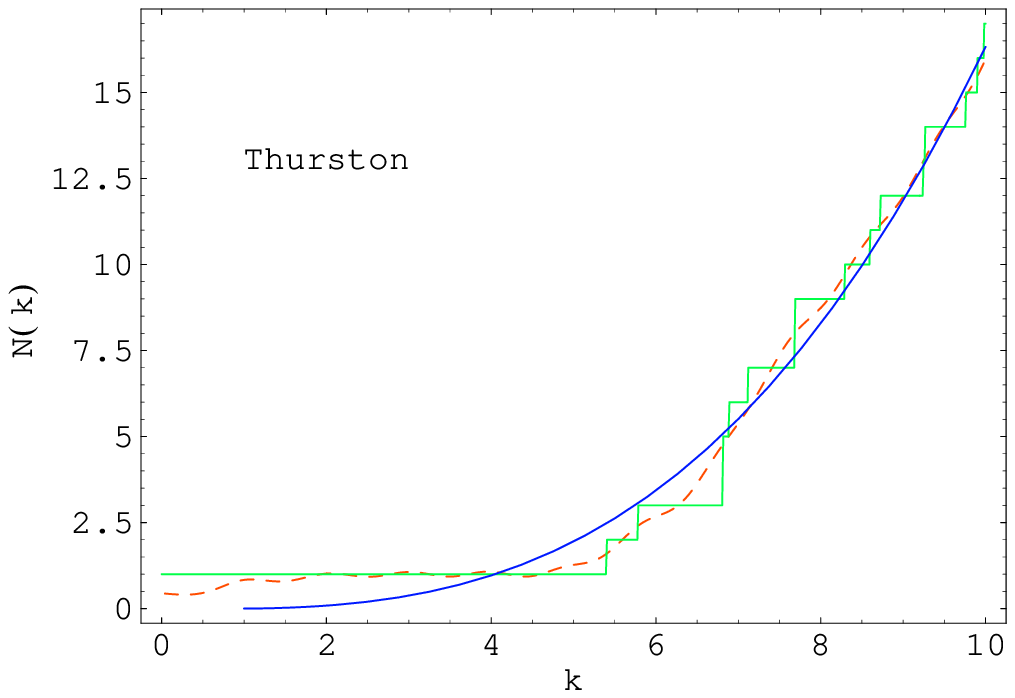,width=8cm}}
\caption{The spectral staircases $N(k)$ for the Weeks manifold 
and the Thurston manifold 
obtained by the DBEM are
compared with the average parts $\hat{N}(k)$ (solid curves), namely, 
Weyl's asymptotic formula (\ref{eq:Weyl}) 
and smoothed spectral staircase (dotted curves) obtained by the 
periodic orbit sum method(POSM)
using all periodic orbits $l<7.0$.} 
\label{fig:ssT}
\EF
\begin{table}
\begin{center}
\setlength{\tabcolsep}{3pt}
\begin{tabular}{cccccc}   
\multicolumn{1}{c}{manifold} &
\multicolumn{1}{c}{volume} &
\multicolumn{1}{c}{$k_1$(DBEM)} &
\multicolumn{1}{c}{$k_1$(POSM)} &
\multicolumn{1}{c}{m($k_1$)} &
\multicolumn{1}{c}{$\Delta k_1/k_1$}  
\\ \hline
m003(-3,1)&0.9427   &5.27  &5.10  &1 &0.03  \\ \hline  
m003(-2,3)&0.9814   &5.40  &5.34  &1 &0.01  \\   \hline  
m007(3,1)&1.0149    &5.29  &5.37  &1 &0.02   \\ \hline  
m003(-4,3)&1.2637   &4.58  &4.31  &2 &0.06     \\ \hline  
m004(6,1)&1.2845    &4.53  &4.35  &1 &0.04   \\ \hline  
m004(1,2)&1.3985    &4.03  &3.93  &1 &0.02 \\ \hline        
m009(4,1)&1.4141    &5.26  &4.84  &2 &0.08 \\ \hline
\end{tabular}
\end{center} 
\label{tab:accuracy}
\caption{The first (non-zero) wavenumbers $k_1=E_1^{1/2}$ are 
calculated using the DBEM and
using the POSM for seven smallest manifolds.
$k_1$'s agree with 
relative accuracy $\Delta k_1 /k_1=0.01-0.08$.
The multiplicity number of the first eigenmode is 
calculated using the DBEM. }
\end{table}
We can see from table II that the first eigenvalues 
calculated by using length spectra
$l<7.0$ for some smallest known 
CH manifolds lie within several per-cent
of those obtained by the DBEM. Note that 
the eigenvalues are also consistent with 
those obtained by the Trefftz method\cite{Cornish1}.
For two examples in which the first non-zero mode is
degenerated, the eigenvalues are much shifted to lower values owing
to the smoothing effect.
From figure \ref{fig:ssT}, one can see that the curves of the 
obtained smoothed spectral stairs cross the ``true'' stairs 
at almost half height.  A slight deviation from the 
average part of the spectral staircase
is caused by the interference of waves each one of which corresponds to
a periodic orbit.
\\
\indent
\section{First eigenvalue and geometrical quantities}
The estimate of the first (non-zero) eigenvalue 
$E_1=k_1^2$ of the Laplace-Beltrami operator 
plays a critical role in describing the global topology and 
geometry of manifolds. A number of estimates of $E_1$ 
for  $n$-dimensional compact 
Riemannian manifolds $M$ 
using diffeomorphism-invariant quantities have been 
proved in mathematical literature.
\\
\indent
First of all, we consider the relation between the first eigenvalue 
$E_1$ and the \ti{diameter} $d$
which is defined as the maximum of the minimum geodesic distance
between two arbitrary points on $M$. Various analytic upper and 
lower bounds of $E_1$ in terms of $d$ have been known. 
 Suppose $M$ with Ricci curvature bounded
below by $-L(L>0)$. Cheng and Zhou proved that $E_1$ satisfies 
\BE
E_1 \ge \max \biggl [ \f{1}{2} \f{\pi^2}{d^2}-\f{1}{4}L,
\sqrt{\f{\pi^4}{d^2}+\f{L^2}{16}}-\f{3}{4}L,
\f{\pi^2}{d^2}\exp \bigl(-C_n \sqrt{Ld^2}/2\bigr) 
\biggr ],
\label{eq:Cheng}
\EE
where $C_n=\max[\sqrt{n-1},\sqrt{2}]$\cite{CZ}. Another lower bound 
has been obtained by Lu\cite{Lu}. Suppose that the Ricci
curvature of $M$ is bounded below as $R_{ab}\ge -K g_{ab},(K\ge 0)$
for a some real number $K$ 
where $g_{ab}$ is the metric tensors of $M$. 
Then $E_1$ satisfies
\BE
E_1 \ge \max \biggl [ \f{\pi^2}{d^2}-K,
\f{8}{d^2}-\f{K}{3},\f{8}{d^2} \exp \biggl ( - \f{d^2 K}{8}
\biggr ), 
\f{8}{d^2}\biggl (1+\f{d}{3}\sqrt{K(n-1)} \biggr) 
\exp\biggl ( - \f{d}{2} \sqrt{K(n-1)} \biggr ) \biggr ].
\label{eq:Lu}
\EE   
As for upper bounds, the following theorem has been proved by 
Cheng\cite{Cheng}. Suppose M with Ricci curvature 
larger than $(n-1)c$, then we have
\BE
E_1\le \tilde{E}_1\bigl ( V_n (c,d/2) \bigr )
\label{eq:Chen}
\EE
where $V_n(c,r)$ denotes a geodesic ball 
with radius $r$ in the $n$-dimensional simply-connected space
with sectional curvature $c$ and $\tilde{E}_1$ is the first Dirichlet
eigenvalue.  
Setting $L\!=\!K\!=\!2,n\!=\!3$, and $c\!=\!-1$, we obtain the upper and 
lower bounds of $E_1$ for CH 3-manifolds. For the upper bound,
(\ref{eq:Chen}) gives a simple relation, $d\le 2 \pi/\nu_1$,
where $\nu_1^2=k_1^2-1$. The physical interpretation is 
clear: the wavelength $\lambda_1\!\equiv \!2 \pi/\nu_1$
of the lowest non-zero mode 
must be larger than the diameter.   
\BF
\centerline{\psfig{figure=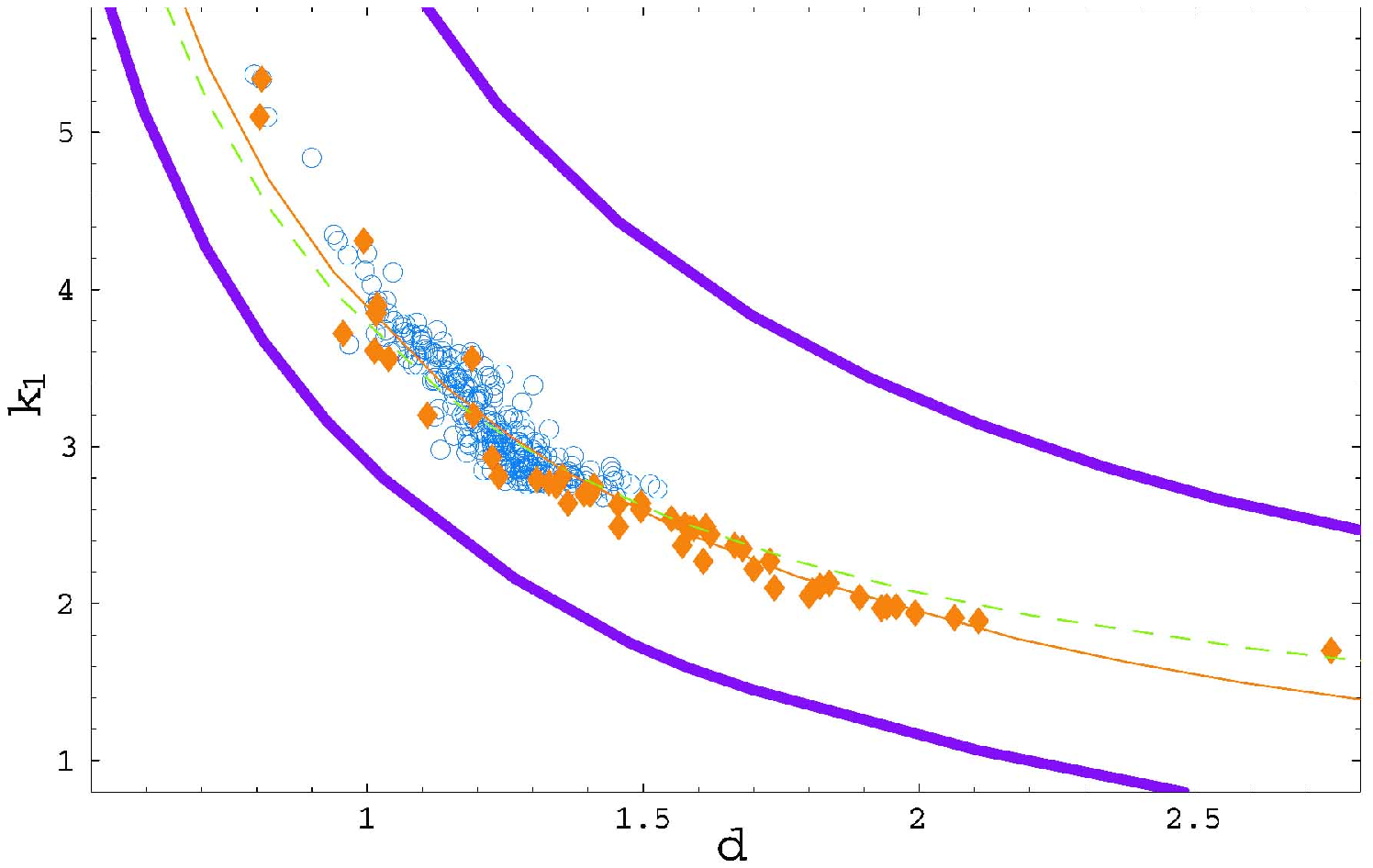,width=13cm}}
\caption{Diameter $d$ versus $k_1$ for 263 examples of CH 3-manifolds
with $l_{min}\!>\!0.3$ and $v\!<\!3$ (group A, circle) and 52 examples
that are obtained by performing Dehn surgeries (51 examples for 
$|p|,|q|<10$ plus 1 example $(p,q)=(16,13)$ group B, diamond) 
on a cusped manifold m003 with 
the best-fit curves for group B corresponding to (\ref{eq:fitA}) 
(dashed curve) assuming $k_1(\textrm{cusp})=1$ 
and (\ref{eq:fitB}) (solid curve) assuming  
$k_1(\textrm{cusp})=0.1$. The fitting curves also agree with 
the computed values for group A.
The upper and lower thick curves denote the
analytic bounds (\ref{eq:Cheng}), (\ref{eq:Lu}) and (\ref{eq:Chen}).  }
\label{fig:dk1}
\EF
\\
\indent
Now, we compare the first eigenvalues  
of 263 examples of CH 3-manifolds with volume less than 3
which have the length of the shortest periodic orbit 
$l_{min}>0.3$
(the Hodgson-Weeks census\cite{HW}) 
and of 45 other examples obtained
by Dehn surgeries ($|p|<17,|q|<14$) 
on a cusped manifold m003\footnote{The Hodgson-Weeks 
census with volume less than 3
also includes 8 manifolds
obtained by Dehn surgeries on m003.} with the analytic
bounds. The diameter of a CH 3-manifold is
given by the supremum of the outradius\footnote{The outradius at 
a basepoint $x$ is equal 
to the minimum radius of the simply-connected 
ball which encloses the 
Dirichlet domain at $x$.}
over all the basepoints,  which has been numerically computed 
using the SnapPea kernel\footnote{I would like to thank J. Weeks
for providing me a code to compute the diameter using the SnapPea kernel.}.
The numerical accuracy is typically $\Delta
d=0.03-0.09$ depending on the topology of the manifold.
\\
\indent
As shown in figure \ref{fig:dk1}, the eigenvalues are well described
by an empirical fitting formula $\lambda_1=\beta d$, or
\BE
k_1=\sqrt{1+\f{4 \pi^2}{\beta^2 d^2}}.
\label{eq:fitA}
\EE
Applying the least square method for the 263 manifolds in the 
Hodgson-Weeks census (group A), and 
53 manifolds obtained by Dehn surgeries on
m003 (group B), the best-fit values $\beta=1.70,1.73$ have been
obtained for each group, respectively.
Note that $\beta\sim1.7$ is slightly larger than the 
values 1.3-1.6 for 12 examples in the previous result 
by Cornish and Spergel\cite{Cornish1}. The deviation from the
fitting formula (\ref{eq:fitA}) is found to be remarkably small
(with one sigma error $\Delta k_1=0.18,0.19$ for each group,
respectively), which implies the existence of much  
sharper bounds. 
\\
\indent
The empirical formula (\ref{eq:fitA}) asserts that no supercurvature 
modes ({\it{i.e.}} $k_1\!<\!1$) exist in the limit 
$d \rightarrow \infty$ where the manifold converges
to the original cusped manifold. However, cusped manifolds may
have some supercurvature modes even for those with small volume. 
Instead of (\ref{eq:fitA}), we consider a generalised empirical formula 
\BE
k_1=\sqrt{(k_1 (\textrm{cusp}))^2+\f{4 \pi^2}{\beta^2 d^2}},
\label{eq:fitB}
\EE
where $k_1 (\textrm{cusp})\le 1$ is the smallest wavenumber for the 
original cusped manifold. Although no supercurvature modes
were observed in this analysis, the non-existence of 
such modes in the limit $d\rightarrow \infty$ was not confirmed
since the
numerical accuracy becomes worse for manifolds with 
small $k_1$ and large $d$.
\\
\indent
Next, we consider the relation between diameter $d$ and volume $v$
of the manifold. Since  diameter is given by the supreme
of the outradius (minimum radius of a sphere which circumscribes the
Dirichlet domain) all over the basepoints, one expect that 
$v$ is estimated as the volume of a sphere (in a hyperbolic space) 
with radius $r$ somewhat
smaller than $d$ if there is no region which resembles the
neibourhood of a cusp (``thin part''
\footnote{A ``thin'' part is defined as
a region where the injectivity radius(a half the minimum length of the
periodic orbit) is short.}) or equivalently $l_{min}$ is 
sufficiently large. 
Suppose that $r=\alpha d$ with $\alpha <1$
then the volume of a sphere  
\BE
v=\pi (\sinh (2 \alpha d)-2 \alpha d)
\label{eq:dvola}
\EE
gives the approximate value of a CH manifold
with diameter $d$. The best fit value for a sample of 79
manifolds with $l_{min}\!>\!0.5$ is $\alpha=0.69$.
If a manifold has a ``thin'' part then 
the relation (\ref{eq:dvola}) is no longer valid
since the diameter becomes too long.  Let $v_c$ be the volume of a
cusped manifold $M_c$ (with only one cusp) and $v(d)$ be the 
volume of a CH manifold $M$ 
obtained by a Dehn surgery on $M_c$. 
In the limit $d \rightarrow \infty$, one can show that 
the following approximation holds (see appendix B):
\BE
v(d)=v_c\biggl (1-\f{\exp(-2(d-d_0))}{\delta+1} \biggr ),
\label{eq:volapp}
\EE
where $\delta$ denotes a ratio of the volume of the complementary part
to that of the ``thin'' part and $d_0$ is the diameter of
the complementary part. For a sample of 41 manifolds with diameter
longer than 1.65 in group B, the best fits are
$\delta=0.0$ and $d_0=0.25$. As shown in figure \ref{fig:dvol},
the volume-diameter relation for CH manifolds with 
large $l_{min}$ is well described
by the fitting formula (\ref{eq:dvola}). As $l_{min}$ 
becomes smaller, or equivalently, $d$ becomes larger, 
a CH manifold $M$ converges to the
original cusped manifold $M_c$ in which (\ref{eq:volapp})
holds.
\BF
\centerline{\psfig{figure=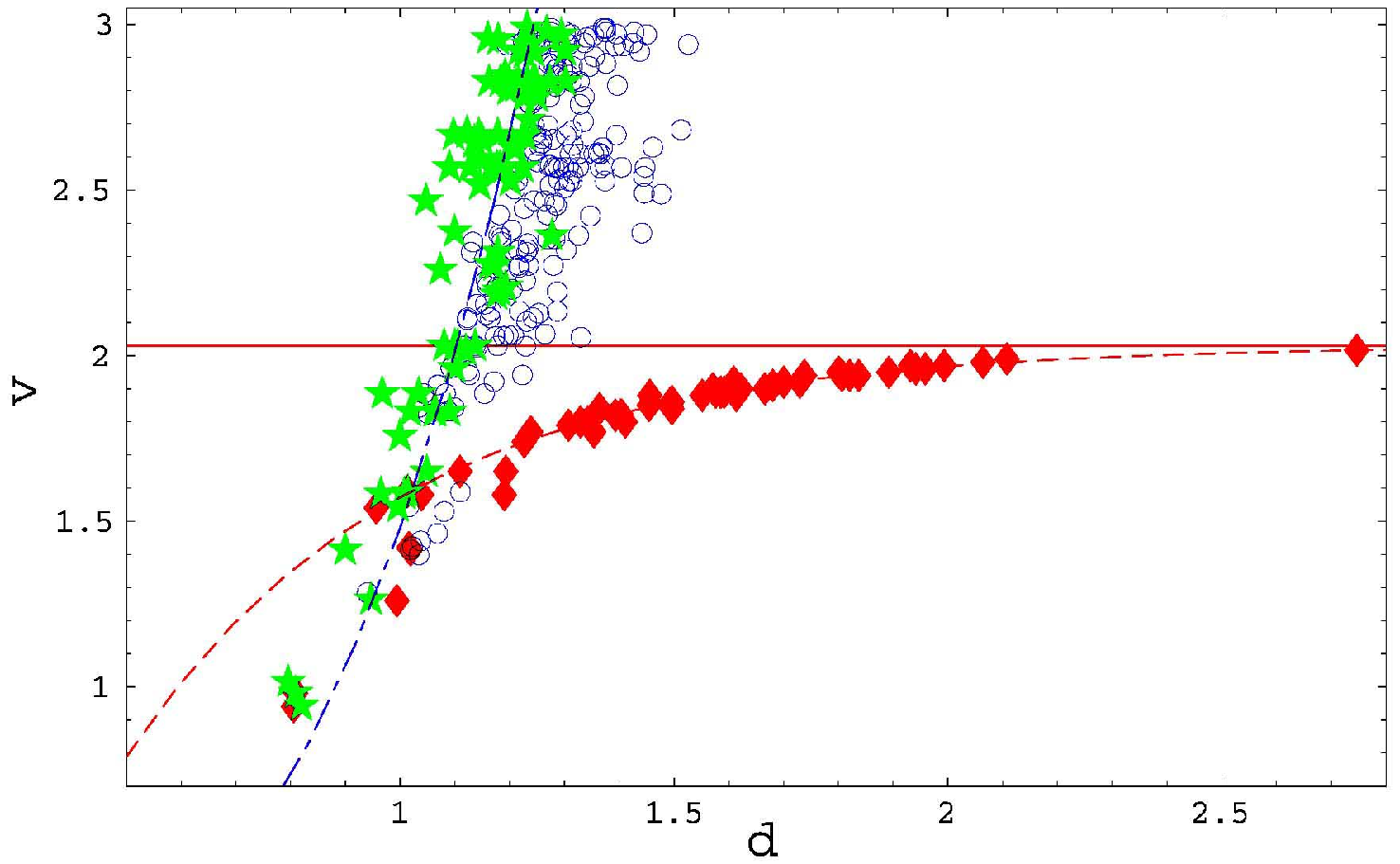,width=13cm}}
\caption{Diameter $d$ versus volume $v$ for 79 manifolds 
($l_{min}\!>\!0.5$)(star), 184 manifolds 
($0.5\!>\!l_{min}\!>\!0.3$)(circle) and 53 manifolds 
obtained by performing Dehn surgeries on m003 (group B, diamond)
with the best-fit curves  
(\ref{eq:dvola}) (dashed-dotted curve) and (\ref{eq:volapp}) 
(dashed curve). 
The solid curve denotes the volume of m003. }

\label{fig:dvol}
\EF
\\
\indent
Finally, we look into the relation between the first eigenvalue and 
the volume. For CH manifolds with sufficiently large
$l_{min}$, (\ref{eq:fitB}) and (\ref{eq:dvola}) give 
\BE
v(k_1)=\pi(\sinh (g(k_1))-g(k_1)),~~~ g(k_1)=\f{4 \pi \alpha}{\beta \sqrt{
(k_1)^2-(k_1(\textrm{cusp}))^2}}.
\label{eq:volk1iso}
\EE
To be consistent with the Weyl's asymptotic formula which is valid for 
$k_1\!>\!>\!1$
\BE
k_1(v)
=\sqrt{\biggl ( \f{9 \pi^2}{v} \biggr )^{\f{2}{3}}+1},
\label{eq:k1Weyl}
\EE
the fitting parameters should satisfy  
$\alpha/\beta=3\cdot 4^{-5/6}\pi^{-2/3}\approx 0.44$ which well
agree with the numerically computed 
values $\alpha/\beta=0.69/(1.6-1.7)=0.41-0.43$ provided that 
$k_1(\textrm{cusp})=1$.  
One can see from figure \ref{fig:vk1} that 
both (\ref{eq:volk1iso}) and (\ref{eq:k1Weyl}) give a good estimate
of the first eigenvalue $k_1^2$
for globally ``slightly anisotropic'' 
manifolds with $v<3$ and $l_{min}>0.5$. 
\BF
\centerline{\psfig{figure=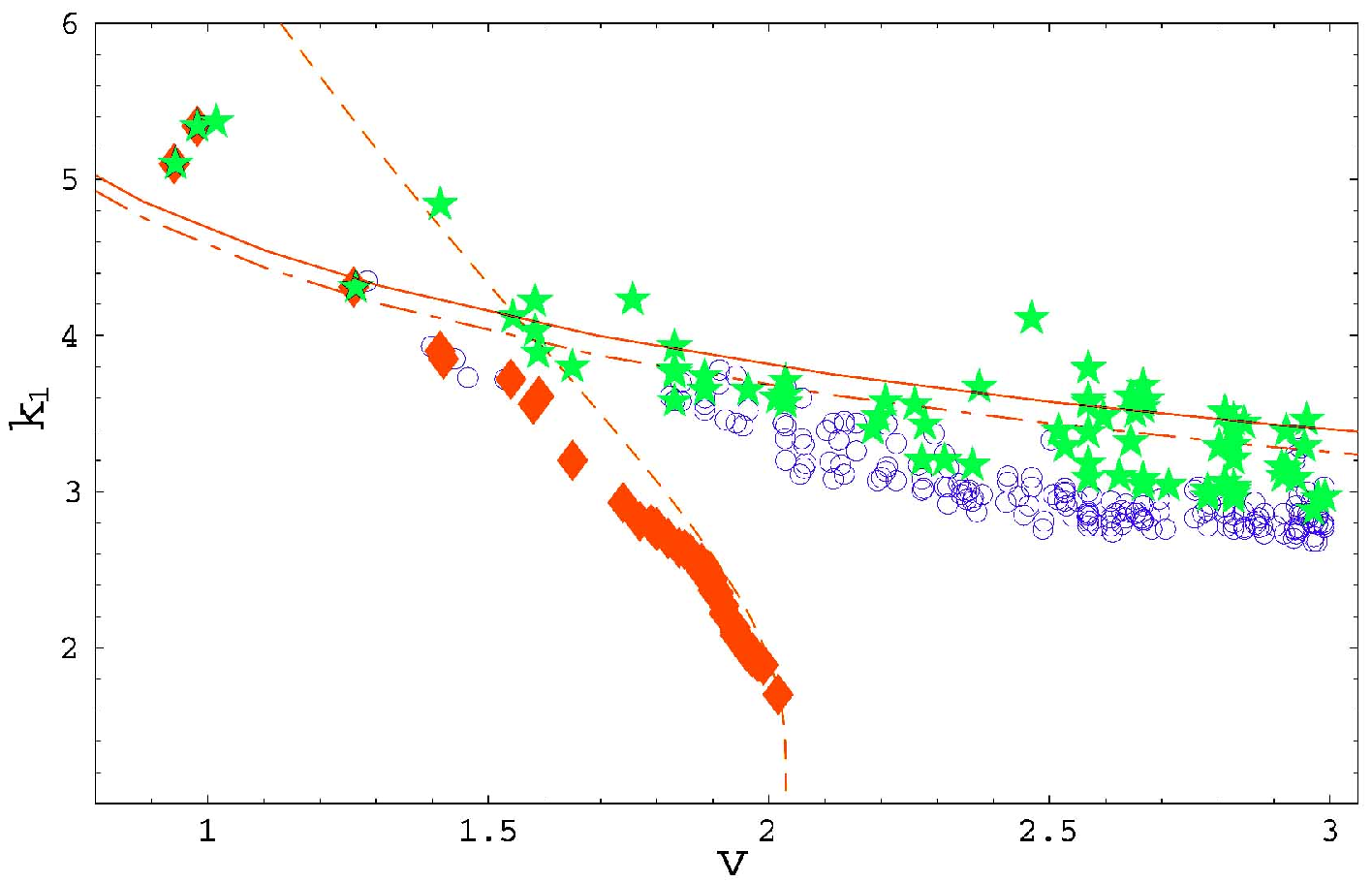,width=13cm}}
\caption{Volume $v$ versus $k_1$ for 79 manifolds 
with $l_{min}\!>\!0.5$ (star), 184 manifolds 
with $0.5\!>\!l_{min}\!>\!0.3$ (circle) and 53 manifolds 
obtained by performing Dehn surgeries on m003 (group B, diamond)
with fitting curves 
(\ref{eq:volk1iso})(solid curve), (\ref{eq:k1Weyl})(dashed-dotted
curve), and (\ref{eq:k1vcusp})(dashed curve) 
where $k_1(\textrm{cusp})=1$ is assumed. 
}
\label{fig:vk1}
\EF
For manifolds with large $d$, $v$ and small $l_{min}$,
(\ref{eq:volk1iso}) and (\ref{eq:k1Weyl}) give
incorrect estimates for $k_1$. In that case, 
the effect of the curvature cannot be negligible. 
In contrast to compact flat spaces,
the volume of a sphere in hyperbolic space
increases exponentially as the radius increases. Assuming 
that the relation (\ref{eq:volk1iso}) holds, 
for sufficiently globally ``isotropic'' CH manifolds (large 
$l_{min}$), \ti{$k_1$ is significantly larger than
that for compact flat spaces with the same volume even if one
assumes that $k_1(\textrm{cusp})\sim 0$}. 
However, for CH manifolds converging to the 
original cusped manifold $M_c$,
the formula (\ref{eq:volk1iso}) has to be modified.
If $l_{min}$ is sufficiently small and $d$ is large while
keeping the volume finite then $M$ has a ``thin'' part similar
to the neibourhood of a cusp. Then
one can use an asymptotic formula 
(\ref{eq:volapp}) instead of (\ref{eq:dvola}). 
\BE
k_1(v)=\sqrt{(k_1(\textrm{cusp}))^2+\f{4 \pi^2}{\beta^2
(d_0-\ln(1-v/v_c)/2)^2}},
\label{eq:k1vcusp}
\EE
where $v_c$ is the volume of $M_c$. 
As shown in figure \ref{fig:devk1minl}, 
$k_1$ shifts to a smaller value for
manifolds with smaller $l_{min}$ which 
have larger $d$. 
\BF
\centerline{\psfig{figure=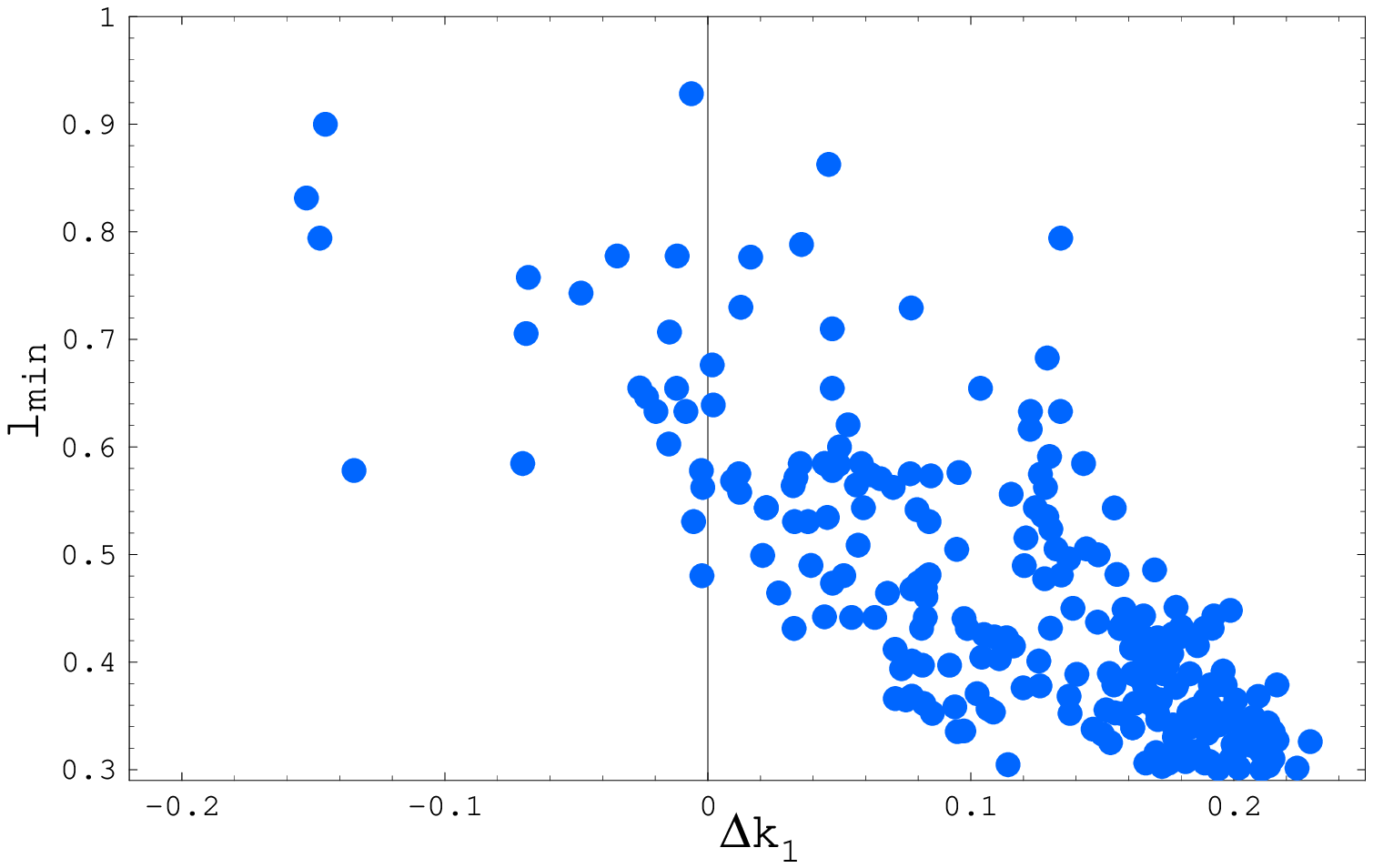,width=9cm}
\psfig{figure=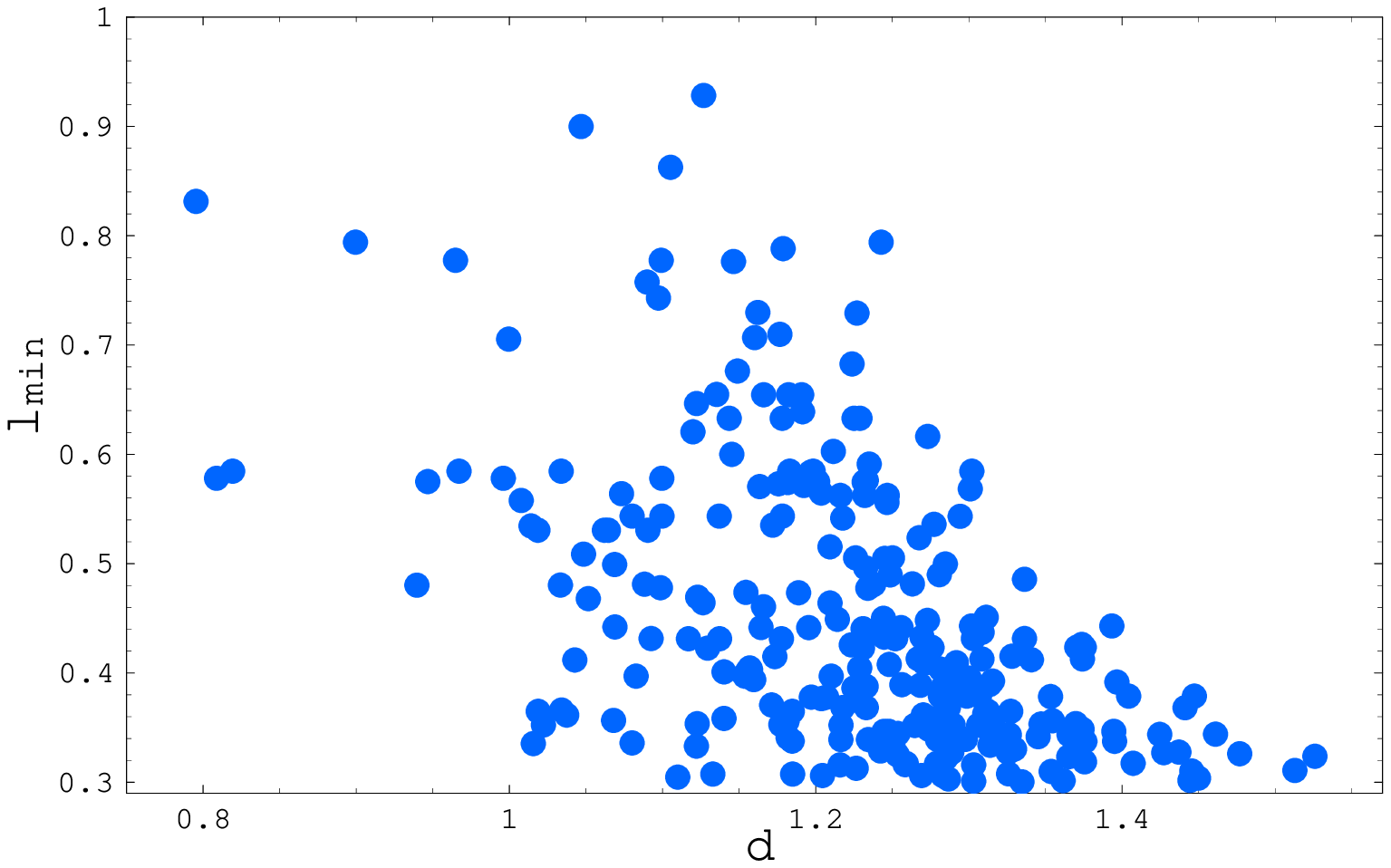,width=9cm}}
\caption{The length of the shortest periodic orbit 
$l_{min}$ versus  
deviation $\Delta k_1$ from the fitting formula 
(\ref{eq:volk1iso}) where $k_1(\textrm{cusp})\!=\!1$ is assumed(left)
 and $l_{min}$ versus  
diameter $d$ (right) for 263 CH manifolds (group A).}
\label{fig:devk1minl}
\EF
In the limit
$M \rightarrow M_c$, the length of 
periodic orbits of $M$ also converges
to that of $M_c$ except for the shortest orbit whose length 
$l_{min}$ goes to zero. From the general Selberg 
trace formula (\ref{eq:GTF}), one can see that the 
wave which corresponds to the shortest orbit has a 
large amplitude $\sim 1/\cosh{l_{min}}$ with a long
wavelength $\sim 2 \pi/l_{min}$. Therefore, the presence
of a very short periodic orbit results in the deviation 
of the energy spectrum
at low-energy region (small $k$) from the asymptotic distribution. 
To confirm this,  the spectral staircase 
using the length spectra of m003(16,13) but the shortest periodic 
orbit is removed has been computed. Note that m003(16,13)
is very similar to the original cusped manifold m003
having $d=2.75$ and $l_{min}=0.0086$. As shown in 
figure \ref{fig:sdev}, the computed spectrum staircase agrees well with
Weyl's asymptotic formula. The computed spectrum may coincide
with the one for a manifold in which the ``thin'' part is cut off. 
\BF
%\centerline{\psfig{figure=Nm003p9q7.eps,width=9.5cm}
%\psfig{figure=Nm003p16q13.eps,width=9.5cm}}
\caption{ 
Spectral staircases for m003(9,7) (dashed-dotted staircase, left)
, m003(16,13) (solid staircase, right) and m003(16,13) without 
the shortest periodic orbit (dotted staircase, right) are shown 
in comparison with the corresponding Weyl's asymptotic
formula (solid curves) which have been numerically computed by
the POSM using all periodic orbits with length $l<7$.  
$N(k)=k$ perfectly fits the staircase of m003(16,13) 
for $k<5$ (thick line, right). }
\label{fig:sdev}
\EF
On the other hand, the spectral staircases for
m003(9,7) and m003(16,13) which have small $l_{min}$
deviate from the average part 
(=Weyl's asymptotic formula(\ref{eq:Weyl})) (figure \ref{fig:sdev}).  
These manifolds have a ``thin'' part which is 
virtually one-dimensional object.  Let us remind that
Weyl's asymptotic formula
for a n-dimensional compact manifold $M$ is given by
\BE
N(k)\sim \f{\omega_n v(M) k^n}{(2 \pi)^n},~~~~k>>1,
\label{eq:Weyln}
\EE
where $\omega_n=2 \pi^{n/2}/(n\Gamma(n/2))$ 
is the volume of the unit disk in the Euclidean 
n-space and $v(M)$ is the volume of $M$. If one assumes that 
the Weyl formula  still holds for small $k$, then 
the spectrum in the low-energy region
which corresponds to fluctuations on large scales for
these manifolds can be approximately 
described by the Weyl formula for $n=1$, $N(k)\propto k/\pi$.
For m003(16,13), it is numerically found that $N(k)=k$ 
provides a good fit for $k<5$.  
In the next section, we will measure the deviation from
the asymptotic distribution
using the low-lying eigenvalue spectra.
\section{Spectral measurement of global anisotropy}
Among many possibilities, we should choose
physically well-motivated quantities for measuring the global 
``anisotropy'' in geometry in terms of eigenvalue spectra.
First, we consider $\zeta$-function which is relevant to 
the cosmological microwave background anisotropy. The 
angular power spectra for CH universes are 
approximately written as\cite{Inoue2}
\BE
C_l\sim \sum_{i=1}^{\infty} \f{F_l(k_i)}{k_i^3},
\EE
where $F_l(k)$ can be approximated as a polynomial function of $k$. 
In order to measure the ``anisotropy'', we will define the following 
parameter,
\BE
\Delta(s)\equiv \zeta(s)/\zeta_w(s)= 
\sum_{i=1}^{\infty} k_i^{-2 s} \bigg / ~
\sum_{i=1}^{\infty} k_{wi}^{-2 s},
\EE
where $k_i^2$ are the eigenvalues of the 
Laplace-Beltrami operator on a CH 3-manifold $M$
with volume $v$
and  $k_{wi}^2$ are the eigenvalues obeying
Weyl's asymptotic formula 
with volume $v$ (\ref{eq:k1Weyl}). Note that the zero-mode $k_0=0$ is not
included in the summation.
Here we only consider 
the case $s>1$ which ensures the convergence of the sum. 
The numerical result shows the clear difference
between the `slightly anisotropic'' manifold m003(-3,1) 
($d\!=\!0.82,l_{min}\!=\!0.58$)
, ``somewhat anisotropic'' manifold m003(9,7) 
($d\!=\!2.11,l_{min}\!=\!0.028$) and ``very anisotropic'' manifold
 m003(16,13) ($d\!=\!2.75,l_{min}\!=\!0.0086$)
(figure \ref{fig:dis}). The presence of the 
fluctuations on large scales in these ``anisotropic'' manifolds
shifts the corresponding $\zeta$-function to a larger value.
For the case in which the shortest periodic orbit is removed from the 
length spectra of m003(16,13), one can see that the spectrum
coincides with that obeying Weyl's asymptotic formula. 
\\
\indent 
Next,  we consider the spectral distance $d_s$ proposed by
Seriu which measures the degree of  semi-classical quantum 
decoherence between two universes having one massless scalar 
field\cite{Seriu},
\BE
d_s[M,\tilde{M}]\equiv\f{1}{2}\sum_{i=1}^\infty \ln \f{1}{2} \biggl (
\f{k_i}{\tilde{k}_i}+\f{\tilde{k}_i}{k_i} \biggr ),
\EE
where $k_i^2$ and $\tilde{k}_i^2$ are the eigenvalues of the
Laplace-Beltrami operator on a compact n-manifold 
$M$ and $\tilde{M}$, respectively. Here we choose eigenvalues
$k_{wi}^2$ as $\tilde{k}_i^2$. In practice we introduce
a cutoff in the summation. It is numerically found that 
the summation converges rapidly for the 3 examples, namely, 
m003(-3,1), m003(9,7) and m003(16,13). The contribution of
the first several terms dominates the summation (figure
\ref{fig:dis}). The result implies that a universe 
having a spatial geometry m003(9,7) or m003(16,13)
semiclassically decoheres with a universe having a spatial geometry 
m003(-3,1) (figure \ref{fig:Dirichlet}).  
\BF
\centerline{\psfig{figure=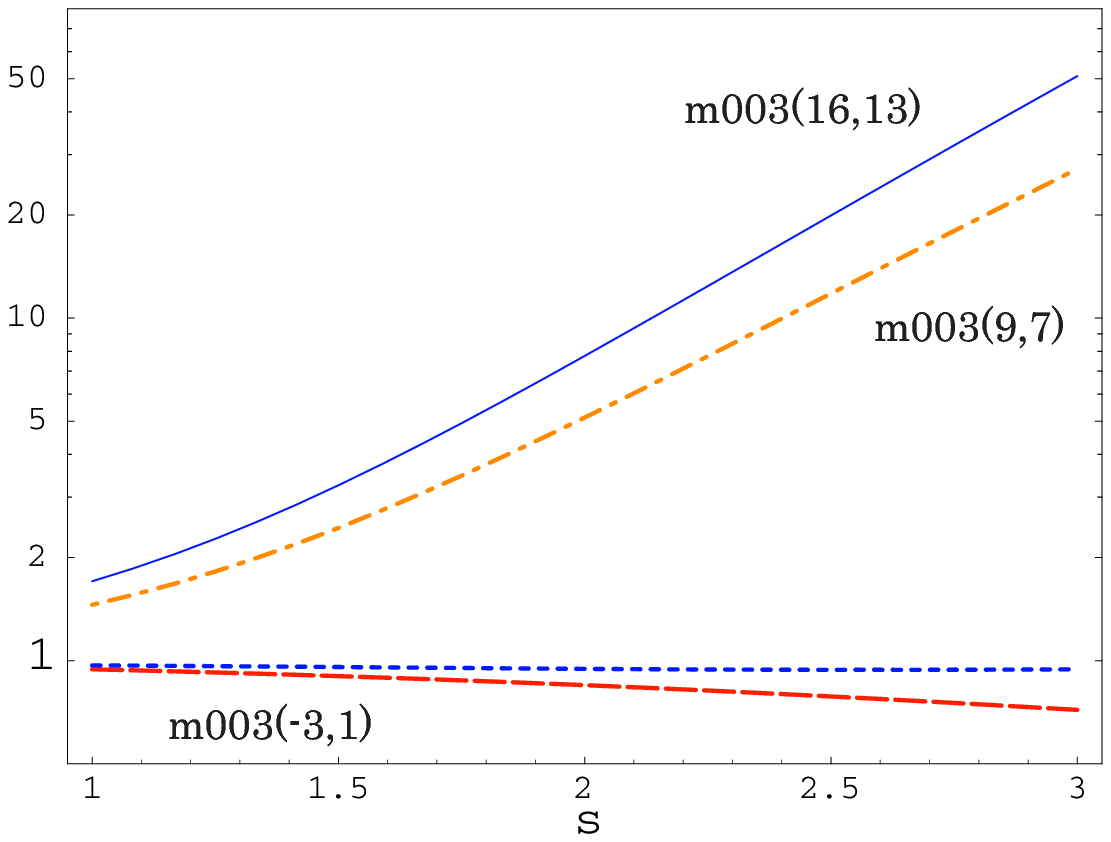,width=9cm}
\psfig{figure=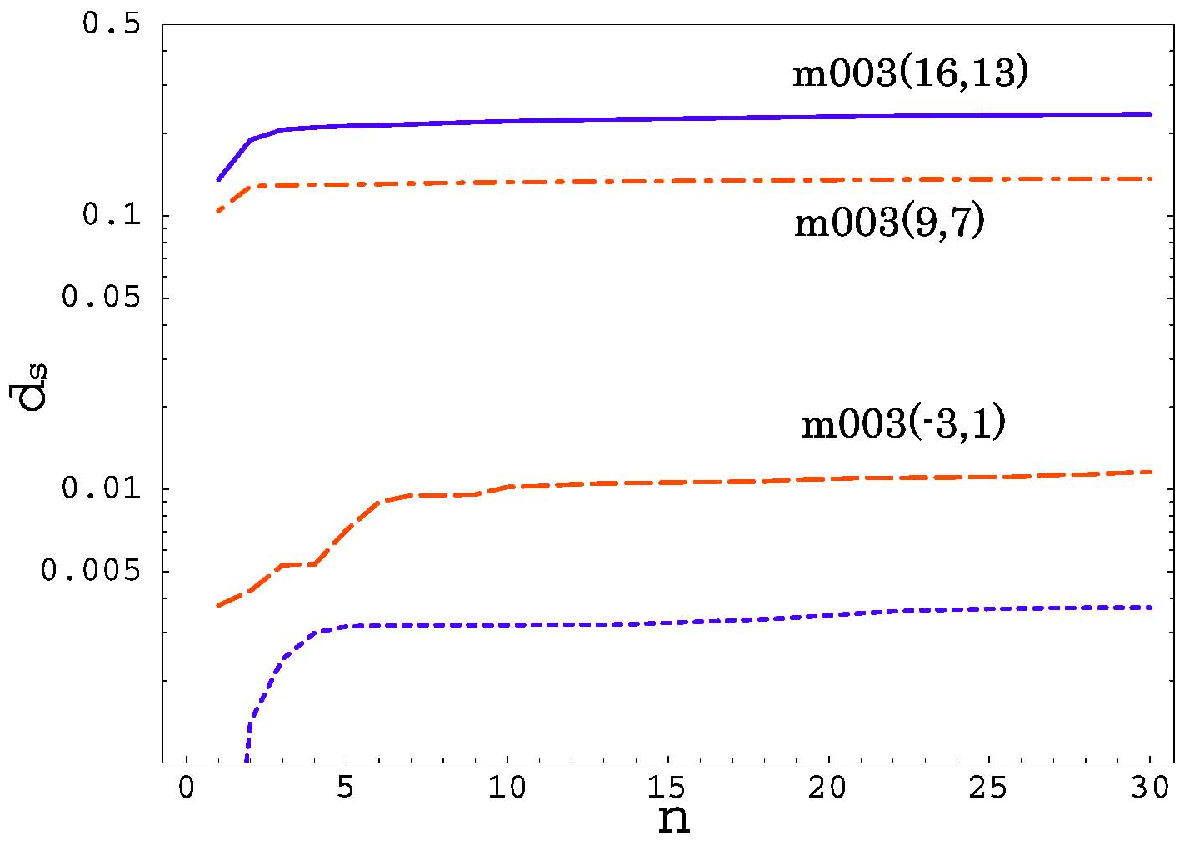,width=9cm}}
\caption{Spectrum measurements of the geometry. The figure
in the left shows  
$\Delta (s)$=$\zeta(s)/\zeta_w(s)$ for 3 examples of CH 3-manifolds
,m003(9,7) (dashed-dotted curve), m003(16,13) (solid curve),
m003(-3,1) (dashed curve) and m003(16,13) where the shortest 
periodic orbit is removed (dotted curve). The figure in the right shows
$d_s(n)=\f{1}{2}\sum_{i=1}^n \ln \f{1}{2} (
k_i/k_{wi}+k_{wi}/k_i )$ for the same examples 
where $n$ is the cutoff number in the summation.}     
\label{fig:dis}
\EF
\BF
\begin{center}
\centerline{\psfig{figure=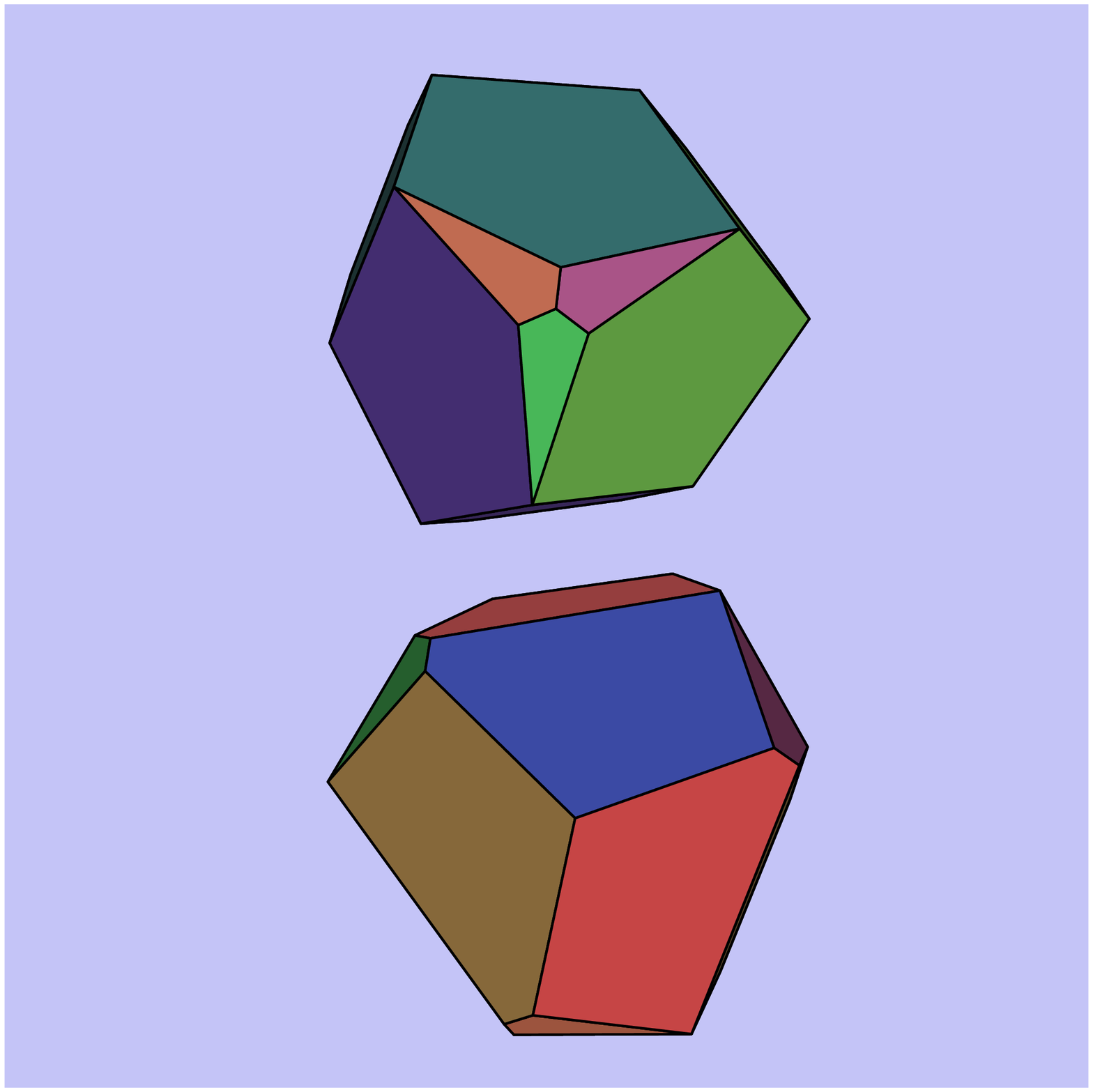,width=8.5cm}
\psfig{figure=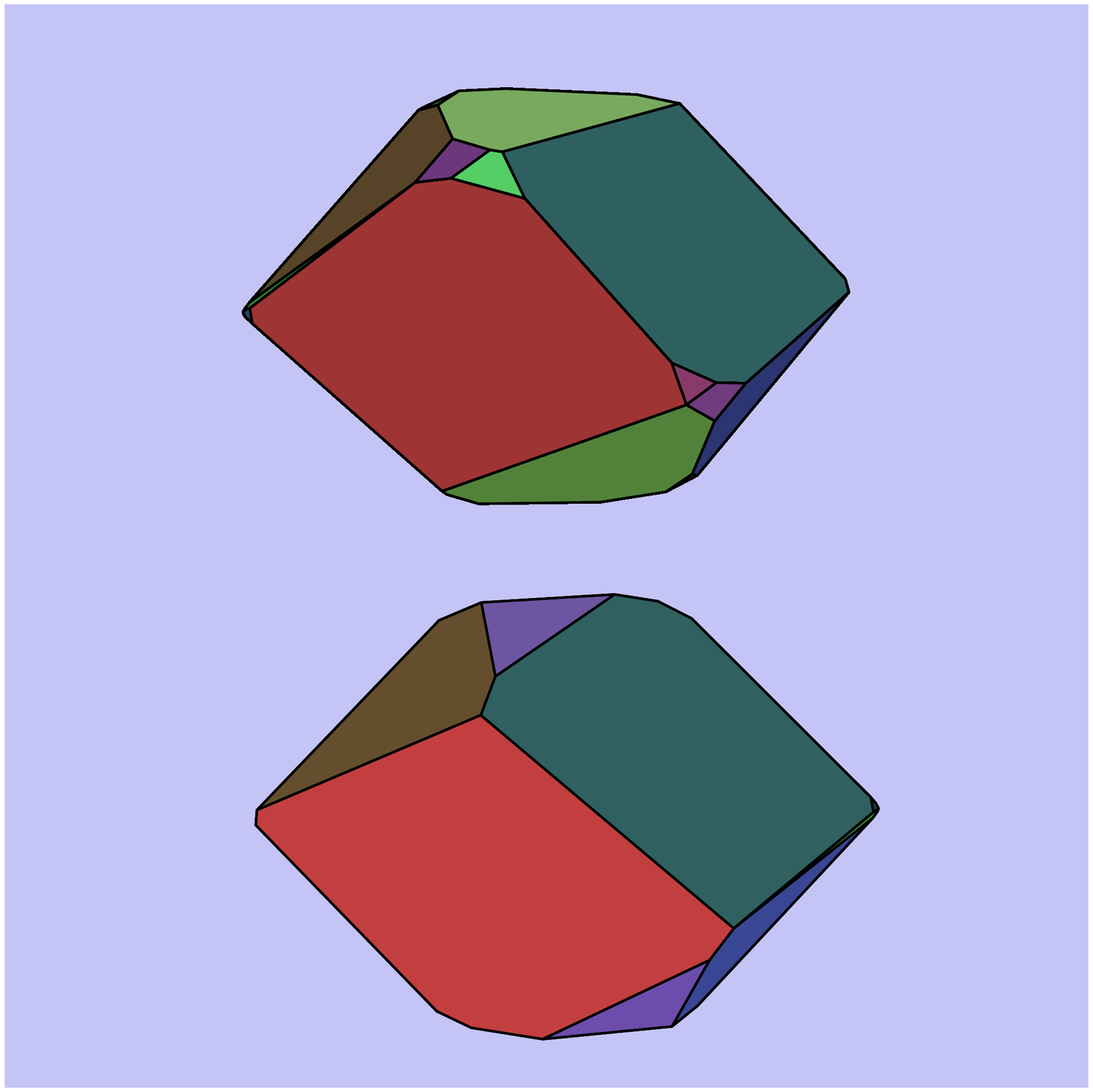,width=8.5cm}}
\caption{Plots of a Dirichlet domain of the 
Weeks manifold m003(-3,1) (left) 
and that of m003(16,13) (right) viewed from two opposite directions 
in the Klein coordinates. The Dirichlet domain of m003(16,13)
is quite similar to that of the original cusped manifold m003
(see figure \ref{fig:fig8}). m003(16,13) has a ``thin'' part
which is similar to the neibourhood of a cusp. The colors on the faces
correspond to the
identification maps. }
\label{fig:Dirichlet}
\end{center}
\EF
\\
\indent
\pagebreak
\section{Summary}
In this paper, the length spectra and low-lying eigenvalue spectra
of the Laplace-Beltrami operator on 
small CH 3-manifolds 
have been numerically investigated. 
CH 3-manifolds are relevant to a various 
kinds of physical systems. For instance, computing the 
CMB anisotropy in CH models is one of the key issue
which have been investigated for several 
models\cite{Bond,Inoue2,Aurich,CS}.
In order to fully understand to what extent CH models
are constrained by the observations, it is necessary to explore 
the property of low-lying eigenvalues and eigenmodes for 
a large number of manifolds (or orbifolds).
\\
\indent
First, The length spectra for a total of 308 CH 3-manifolds 
(volume less than 3)
have been successfully computed using the SnapPea
kernel up to the length $l=7.0$.  
The asymptotic behavior in the classical staircase 
is found to be consistent with the known analytical 
formula which does not depend on the topology
or symmetry of the manifold. Regarding the symmetry of 
the length spectra, 
it is well known that the arithmetic structure
breaks a generic feature:
the locally averaged multiplicity number grows exponentially
as $l$ is increased. 
However, the exponential behavior in the multiplicity
number which has been noticed for a non-arithmetic 3-orbifold
has been confirmed for the smallest five examples of 
non-arithmetic 3-manifolds, though the rate is smaller than
that for the arithmetic manifolds. 
This may be related to the ``hidden symmetry'' which is
the symmetry of finite sheeted covers of the manifold (which can be
tessellated by the copies of the fundamental domain of the manifold) 
but not of the manifold itself. 
\\
\indent
Next, the trace formula has been applied to these 3-manifolds for
computing low-lying eigenvalues using the length spectra (POSM).
Consistency with those obtained by the DBEM
has been confirmed for several manifolds.
It is numerically 
found that these manifolds do not have any supercurvature modes.
It seems that the manifolds which supports supercurvature modes
are either having large volume or large diameter. In order to
confirm this, further investigation of eigenvalues for cusped
manifolds is necessary.  
\\
\indent
Thirdly, the first eigenvalues are compared to a various
diffeomorphism-invariant quantities, namely, diameter, volume
and the shortest length of the periodic orbits. The 
numerical results imply the existence of much shaper bounds
for the first eigenvalues in terms of diameter.  
Some fitting formulae have been introduced and
their validity has been checked. 
We have seen that 
CH 3-manifolds can be roughly divided into two categories:
``slightly anisotropic'' and ``almost anisotropic'' ones.
The former has not any very short periodic orbits while the 
latter has. For example, manifolds which are very similar 
to the original cusped manifold are belonging to the latter
category. It is found that the  
deviation of the spectrum from the Weyl asymptotic formula 
for these manifolds is conspicuous even for manifolds 
with small volume.
\\
\indent
Finally, the global ``anisotropy'' in the spatial 
geometry has been measured by
$\zeta$-function and the spectral distance
for 3 examples of CH 3-manifolds. These measurements
give a clear indication 
of the presence of a ``thin'' part in the manifold in terms 
of eigenvalue spectra. In other words, 
the physical quantities (the angular power spectra in the CMB or
decoherence between two universes) are greatly affected 
by the globally anisotropic structure in the spatial geometry.  
\vspace{0.5cm}
\section*{Acknowledgments}
I would like to thank Jeff Weeks for his extensive 
advice on the use of SnapPea and excellent explanation on 
geometry and topology of CH spaces and Ralph Aurich 
for his informative comments 
on the trace formula. I would also like to thank Ian Agol
for pointing out some mistakes in the manuscript.
The numerical computation in this work was carried out  
at the Data Processing Center in Kyoto University and 
Yukawa Institute Computer Facility. 
K.T. Inoue is supported by JSPS Research Fellowships 
for Young Scientists, and this work is supported partially by 
Grant-in-Aid for Scientific Research Fund (No.9809834). 

%======================================%
%<<<<<<<<<<<< Bibliography >>>>>>>>>>>>%
%======================================% 

%==================================%
%<<<<<<<<<<<< Appendix >>>>>>>>>>>>%
%==================================% 
\pagebreak
\appendix
\section{} 

 For a given Dirichlet domain $D$, SnapPea computes
\\
\\
1. Neighboring copies of $D$ (tiles)   
   recursively and stores the corresponding 
   elements $g$ of the discrete isometry group $\Gamma$.
   If it has already been found out, 
   it is discarded. The computation proceeds
   until for all the neighborhoods of $gD$,
   $d(x,hx)> \cosh^{-1} (\cosh R \cosh l/2)$ satisfies
   where $hD'$s are neighborhoods of $gD$($R$ is the spine radius 
   and $h$ is an element of $\Gamma$).
   Since there is no $g$ other than identity 
   where all $h'$s satisfy $d(x,hx)\ge d(x,gx)$,
   this algorithm will not miss any tiles $gD$
   where  $d(x,gx)<2 \cosh^{-1} (\cosh R \cosh l/2)$. 
\\
\\
2. A list of geodesics 
   for all $g's=\{g\}$ where 
   1:the real part of length is not zero and less than $l$;
   2:the distance from $x$ to the geodesic is at most $R$.
\\
\\   
3. The conjugacy class $g'=h g h^{-1}$ or its inverse for 
   each $g$ where $h$ is an element of $\{g\}$. If an identical 
   complex length is found, the complex length which corresponds to
   $g'$ is omitted from the list of geodesics. If the complex
   length of $g$ is conjugate to that of $g^{-1}$, the geodesic
   is topologically a mirrored interval, otherwise it is a circle.
\\
\\
4. Multiplicity of geodesics.
   If a pair of geodesics with two complex lengths being 
   identical within an error range is found, the multiplicity
   number is increased by one. 

\section{}
 
Suppose a CH manifold $M$
which resembles the original cusped manifold $M_c$ with one cusp.
Let us divide $M_c$ into two parts, the neighbourhood
of a cusped point $K_c$, and the complementary part $K_{c0}$.
Similarly, one can divide $M$ into $K$ and $K_{0}$ where
$K_{c0}\approx K_0$ and $K$ corresponds to a ``thin'' part.
Since the neighbourhood of a cusp is represented as a ``chimney''(but
having infinite length)
in the upper half space coordinates $(x_1,x_2,x_3)$, 
$K$ can be well approximated by
an elongated box defined by ($-\Delta x/2 \le x_1 \le \Delta x/2,    
-\Delta x/2 \le x_2 \le \Delta x/2, x_{30}\le x_3 \le x_{31}$).
Then the physical length(=diameter) $\tilde d$ of $K$
in the direction $x_3$ is given by
$\tilde{d}=\ln (x_{31}/x_{30})$. On the other hand, the volume 
$\tilde{v}$ of $K$ satisfies
\BE
\tilde{v}=\f{(\Delta x)^2}{2}\biggl( \f{1}{(x_{30})^2}
-\f{1}{(x_{31})^2} \biggr ),
\EE
which gives the ratio of the volume 
$\tilde{v}$ to the volume of $\tilde{v}_\infty$ of $K_c$,
$\tilde{v}/\tilde{v}_\infty=1-(x_{30})^2/(x_{31})^2
=1-\exp(-2 \tilde{d})$. If we approximate the diameter $d$
of $M$ as $d=\tilde{d}+d_0$ where $d_0$ is the diameter of $K_0$,
then we finally have the ratio of the volume $v=v_0+\tilde{v}$ of $M$ to
the volume $v_c=v_0+\tilde{v}_\infty$ of $M_c$,
\BE
\f{v}{v_c}=1-\f{\exp (-2(d-d_o))}{\delta+1},~~~
\delta\equiv v_0/\tilde{v}_\infty
\EE
where $v_0$ denotes the volume of $K_0$.

\BF
\centerline{\psfig{figure=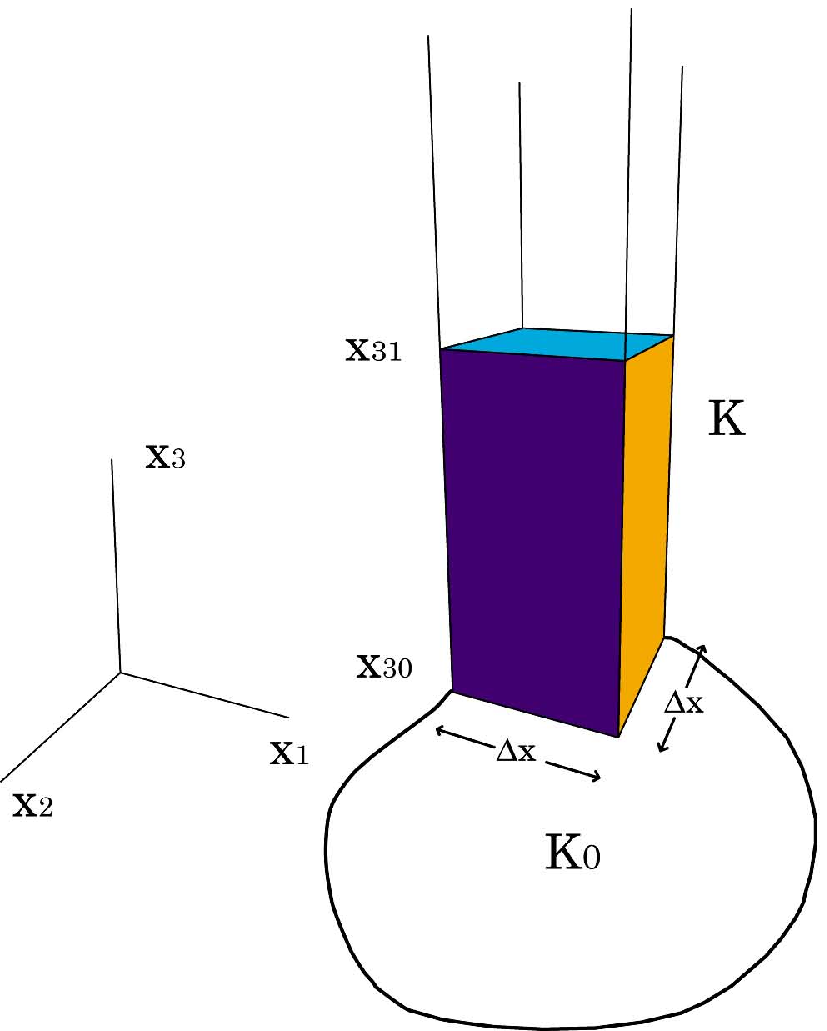,width=8cm}}
\caption{Estimate of the volume of a ``thin'' part $K$ in a manifold
$M$ in the upper half space.}
\label{fig:appb}
\EF

\end{document}